\documentclass{jfm}
\pdfoutput=1
\usepackage{graphicx}
\usepackage{natbib}
\usepackage{hyperref}
\hypersetup{
    colorlinks = true,
    linkcolor = blue,
    urlcolor   = blue,
    citecolor  = black,
}
\usepackage{newtxtext}
\usepackage{newtxmath}
\usepackage{tabularx}
\usepackage{amsmath}
\usepackage{bm}      
\usepackage[geometry]{ifsym}
\usepackage{subcaption}
\usepackage{comment}

\usepackage{xcolor}

\newcommand{\Sto}{\mathrm{St}_\Gamma}
\newcommand{\Atw}{\mathrm{At}}
\colorlet{revision}{black}
\colorlet{revision2}{black}

\title{Instability of a dusty vortex}

\shortauthor{S. Shuai, D. J. Dhas, A. Roy, and M. H. Kasbaoui}

\author{
  Shuai Shuai \aff{1} and
  Darish Jeswin Dhas \aff{2} and
  Anubhab Roy\aff{2} and
  M. Houssem Kasbaoui \aff{1}\corresp{\email{houssem.kasbaoui@asu.edu}}
}

\affiliation{
  \aff{1}School for Engineering of Matter, Transport and Energy, Arizona State University, Tempe, AZ 85281, USA.
  \aff{2}Department of Applied Mechanics, Indian Institute of Technology Madras, Chennai, 600036, India.
}
\begin{document}

\maketitle

\begin{abstract}
    We investigate the effect of inertial particles dispersed in a circular patch of finite radius on the stability of a two-dimensional Rankine vortex in semi-dilute dusty flows. Unlike the particle-free case where no unstable modes exist, we show that the feedback force from the particles triggers a novel instability. The mechanisms driving the instability are characterized using linear stability analysis for weakly inertial particles and further validated against Eulerian-Lagrangian simulations. We show that the particle-laden vortex is always unstable if the mass loading $M>0$. Surprisingly, even non-inertial particles destabilize the vortex by a mechanism analogous to the centrifugal Rayleigh-Taylor instability in radially stratified vortex with density jump. We identify a critical mass loading above which an eigenmode $m$ becomes unstable. This critical mass loading drops to zero as m increases. When particles are inertial, modes that fall below the critical mass loading become unstable, whereas, modes above it remain unstable but with lower growth rates compared to the non-inertial case. Comparison with Eulerian-Lagrangian simulations shows that growth rates computed from simulations match well the theoretical predictions. Past the linear stage, we observe the emergence of high-wavenumber modes that turn into spiraling arms of concentrated particles emanating out of the core, while regions of particle-free flow are sucked inward. The vorticity field displays similar pattern which lead to the breakdown of the initial Rankine structure. This novel instability for a dusty vortex highlights how the feedback force from the disperse phase can induce the breakdown of an otherwise resilient vortical structure.
\end{abstract}


\section{Introduction} \label{sec:Introduction}
 
Vortical flows with heavy dispersed particles occur in diverse scenarios, both of natural and engineering origins \citep{balachandar2010turbulent,guazzelli2011physical}. To cite a few examples, they include atmospheric funnel-type vortices (such as `dust devil')\citep{bluesteinDopplerRadarObservations2004}, centrifugal separation devices, swirling biomass combustors, and aircraft trailing vortices with condensation droplets \citep{paoliContrailModelingSimulation2016,paoliLargeeddySimulationTurbulent2008}. The rudimentary understanding of the process is that particles that are denser than the surrounding fluid drift away from intense vortical regions at a rate controlled by particle inertia. However, even in elementary vortical flows such as columnar vortices, the dispersion of inertial particles (micron-sized solid spheres, or liquid droplets, sufficiently small to remain spherical under the action of surface tension) is difficult to predict accurately. This uncertainty is due to models and simulations often omitting the particle feedback force. Yet, the latter may be significant even in dilute conditions (particle volume fractions is $10^{-5}<\langle \phi_p\rangle<10^{-3}$) provided that the density ratio is high ($\rho_p/\rho_f=O(1000)$), e.g. when the carrier fluid is gas. In this study, we shall show that particle feedback arising from two-way momentum coupling activates a novel vortex instability, and this instability is controlled by particle inertia and vortex structure.
    
To discuss the stability properties in a vortex flow, we adopt the Rankine vortex as a typical vortex tube, i.e., one with uniform vorticity within the core and zero outside \citep{saffman1995vortex}. Many measurements of atmospheric phenomena show that the Rankine vortex is suitable for describing velocity structure of naturally occuring vortical flows. Doppler radar measurements prove that tornados are well approximated by the Rankine vortex profile \citep{bluesteinMobileDopplerRadar2003}. A mesocyclone is another example of atmospheric flow that is well approximated by a Rankine vortex with an inner core of solid rotation as large as  5km \citep{brownImprovedDetectionSevere2005}. 

The simple velocity profile of the Rankine vortex makes stability investigations amenable to analytical treatments, thus, often  {\color{revision}helping} in better identifying the underlying physics. Lord Kelvin \citep{thomsonXXIVVibrationsColumnar1880, lambHydrodynamics1993} analyzed the linear stability of an incompressible Rankine vortex. He showed that for 2D perturbations characterized by mode number $m$, only one real eigenvalue exists, indicating that the perturbation will propagate with no growth or decay, corresponding to neutral stability. Besides these neutrally stable discrete modes, a Rankine vortex also supports  {\color{revision} a continuous spectrum of modes} whose combination may lead to algebraic growth for short times \citep{roy2014linearized}. The inclusion of additional physics can destabilize a vortex column. For instance, acoustic waves radiating to infinity can destabilize a Rankine vortex by extracting energy from the vortex core \citep{broadbent1979acoustic}. Analogous destabilization mechanisms have been shown to exist for shallow water flows \citep{ford1994instability} and stratified flows \citep{schecter2004damping,le2009radiative}, where outgoing waves draw energy from the mean flow. Recently a new instability mechanism has been found for a Rankine vortex in dilute polymer solutions \citep{roy2022inertio}. The instability occurs due to the resonant interaction of a  pair of elastic shear waves aided by the differential convection of the irrotational shearing flow outside the vortex core.
Compressibility effects may also destabilize a Rankine vortex as shown by \citet{sozouAdiabaticPerturbationsUnbounded1987}.

The configuration of a radially stratified vortex has special relevance, since, as we shall discuss in detail later, it bears analogy with a particle-laden vortex in the limit of zero particle inertia. If density increases monotonically with radius, \citet{fungStabilitySwirlingFlows1975} showed that the flow is stable to both axisymmetric and non-axisymmetric modes. In contrast, a heavy-cored vortex may become unstable, as shown by \citet{sippStabilityVortexHeavy2005a}, through two mechanisms, one due to a centrifugal instability involving short-axial wavelength modes, and the second due to a Rayleigh-Taylor instability involving two-dimensional modes. Similarly, \citet{jolyRayleighTaylorInstability2005} found that heavy-cored two-dimensional vortices are subject to a Rayleigh-Taylor instability. They attributed the basic mechanism to baroclinic vorticity generation caused by the misalignment between density gradient and centripetal acceleration. Using numerical simulations, \citet{jolyRayleighTaylorInstability2005}  showed that the unstable modes lead to the formation of spiraling arms in the density and vorticity fields that roll-up eventually as nonlinear effects emerge. \citet{dixitStabilityVortexRadial2011} analyzed the case of a radially stratified vortex with a density jump. They showed that both heavy-cored and light-cored vortices could be unstable if the density jump is sufficiently steep. They proposed that the instability is triggered by the wave-interaction mechanism between the Kelvin wave located at the vortex core edge and counter-propagating internal waves caused by the centrifugal force (see \citet{carpenter2011instability,balmforth2012dynamics} for discussions on instabilities originating from interaction between waves riding on vorticity/density jumps). When the vorticity and density jumps are sufficiently separated, they act as ``independent'' neutral modes; the two interfacial displacements are in phase with each other and out of phase with the radial velocity. Thus, the radial velocity neither aid nor retard the interfacial displacement. When the jumps are located close to each other, the interfacial displacements are out of phase, and the radial velocity reinforces the interfacial displacement at the density jump resulting in a growing mode. \citet{dixitStabilityVortexRadial2011} also found that a heavy-cored vortex is stabilized when the density jump is placed in a region immediately outside the vortex core. Our analysis in section \S \ref{sec:Sto} will reveal that, when we neglect particle inertia but account for non-zero mass loading, the system analyzed in the present study mimics a non-Bousinessq fluid. Thus, the observed instability has an interesting analogy with the physical mechanisms elucidated in the investigations mentioned above.
    
When inertial particles heavier than the carrier fluid disperse in the flow, they tend to migrate from regions of high rotation rate to regions of high strain rate. This mechanism known, {\color{revision}as preferential concentration \citep{squiresPreferentialConcentrationParticles1991, wangSettlingVelocityConcentration1993}}, results from a slip velocity between particles and flow, which increases with increasing particle inertia. Preferential concentrations leads to the clustering of inertial particles around vortical cores.

Although particle dispersion in vortical flows exhibits rich dynamics, the disperse phase does not alter the fundamental stability of the flow unless two-way coupling is considered. Recently there has been a strong interest in the modulation of flow instabilities due to the feedback force from small, heavy particles. \cite{magnani2021inertial} considered the effect of particle inertia in two-way coupled dusty Rayleigh-Taylor turbulence. They observed that the system behaves similarly to an equivalent denser fluid for low inertia particles. When particle inertia increases, turbulent mixing gets delayed. The non-monotonic role of the disperse phase was also observed by \cite{sozza2022instability} in their stability study of two-way coupled dusty Kolmogorov flow. When particle inertia is weak, the instability is enhanced. However, for larger Stokes number, particles can both stabilize or destabilize the Kolmogorov flow, with a non-monotonic dependence on the mass fraction. Inertial particles can also trigger instabilities that would not otherwise exist in particle-free flows. \citet{kasbaouiPreferentialConcentrationDriven2015} showed that the interplay between preferential concentration and gravitation settling triggers a non-modal instability in simple shear flow. This instability may bootstrap additional Rayleigh-Taylor and particle-trajectory crossing instabilities \citep{kasbaouiClusteringEulerEuler2019} and plays a role in the attenuation of turbulent fluctuations in particle-laden homogeneously sheared turbulence \citep{kasbaouiRapidDistortionTwoway2019,kasbaouiTurbulenceModulationSettling2019}.

In the context of vortical flows, few studies considered the impact of two-way coupling on the stability of such flows. \citet{marshallParticleDispersionTurbulent2005b} studied the effect of turbulence on dispersed particles in a vortex column under one-way coupling, i.e., neglecting the particle feedback force. He showed that inertial particles initially inside the vortex core are driven out, forming concentrated ring-like structures. Turbulence breaks these structures into smaller sections. \citet{ravichandranCausticsClusteringVicinity2015a} studied the formation of particle clusters around isolated and ensembles of two-dimensional vortices in one-way coupling. They showed that particles initially within a critical radius, that depends on the vortex circulation and particle response time, form caustics, thus, potentially leading to higher collision rates. When two-way coupling is considered, the dispersion of inertial particles may lead to a modulation of the carrier flow. \citet{druzhininConcentrationWavesFlow1994} showed that the outward ejection of inertial particles from a particle-rich vortex core causes the attenuation of vorticity in the core. Considering two-dimensional axisymmetric perturbations, he showed that the particles form a ring-shaped cluster that expands outwardly, akin to a concentration wave, at a rate controlled by Stokes number $\Sto=\tau_p/\tau_f$, where $\tau_p$ is particle response time, and $\tau_f$ is fluid characteristic timescale. These prior observations were corroborated recently by \citet{shuaiAcceleratedDecayLambOseen2022} in two-way coupled Eulerian-Lagrangian simulations of particle-laden Lamb-Oseen vortex. The authors found that particle rings grow on a timescale $\tau_c=\tau_f/\Sto$, and proposed an expression for their expansion rate. Similar to \citep{druzhininConcentrationWavesFlow1994}, \citet{shuaiAcceleratedDecayLambOseen2022}  observed that particle-feedback on the fluid drives a faster decay of the vortex tube. Further, the observation of naturally emerging azimuthal perturbations in the vorticity field suggests that an instability activated by two-way coupling may be at play. These observations motivate the present study on the stability of a particle-laden vortex.



This paper is organized as follows. Section \S \ref{sec:evidence} presents evidence of instability in a two-way coupled particle-laden vortex. The governing equation and linear stability analysis for small inertia particles are presented in \S \ref{sec:lsa}. In \S \ref{sec:lsa_EL_comp}, we compare the analytical solutions with the results from Eulerian-Lagrangian simulations of the two-way coupled system. Concluding remarks are given in \S \ref{sec:conclusion}.

\section{Evidence of instability in a two-way coupled particle-laden vortex}
\label{sec:evidence}

In this section, we show that the modulation of a prototypical vortex, the Rankine vortex here, is driven by an instability activated by two-way coupling. We illustrate this behavior in a sample flow using Eulerian-Lagrangian simulations.

\subsection{Eulerian-Lagrangian method}
The Eulerian-Lagrangian simulations presented in this work are based on the volume-filtered formulation \citep{andersonFluidMechanicalDescription1967,jacksonDynamicsFluidizedParticles2000,capecelatroEulerLagrangeStrategy2013}. In this formulation, the fluid phase is treated in the Eulerian frame, whereas the particle phase is treated in the Lagrangian frame, i.e., each particle is individually tracked.

The mass and momentum conservation equations for the carrier phase in the semi-dilute regime, are given by the incompressible Navier-Stokes equations,
\begin{eqnarray}
\nabla\cdot \mathbf{u}_f &=&0, \label{eq:2.1}\\
\rho_f\left( \frac{\partial \mathbf{u}_f}{\partial t}+\nabla\cdot \left(\mathbf{u}_f \mathbf{u}_f\right) \right)&=&-\nabla p+\mu_f\nabla ^2 \mathbf{u}_f+\mathbf{F}_p,  \label{eq:2.2}
\end{eqnarray}
where $\mathbf{u}_f$ is the fluid velocity, $p$ is the pressure, $\rho_f$ is the fluid density, and $\mu_f$ is the fluid viscosity. The term $\mathbf{F}_p$ represents momentum exchange between particles and fluid, which is expressed as
\begin{equation}
\mathbf{F}_p=  -\phi_p \nabla\cdot \tau|_p +\rho_p\phi_p \frac{(\mathbf{u}_p-\mathbf{u}_{f|p})}{\tau_p},\label{eq:coupling}
\end{equation}
where $\tau=-p\mathcal{I}+\mu (\nabla u_f+\nabla u_f^T)/2$ is the total fluid stresses, $\phi_p$ is the particle volume fraction, $\mathbf{u}_p$ is the Eulerian particle velocity, and  $(\cdot)_{|p}$ indicates fluid quantities at the particle locations. The first term in (\ref{eq:coupling}) is the stresses exerted by the undisturbed flow at the particle location. The second term is the stresses caused by the presence of particles which are represented using Stokes drag. When the density ratio is large ($\rho_p/\rho_f\gg 1$), as is the case presently, Stokes drag dominates the momentum exchange.

From a scaling analysis of the drag force in (\ref{eq:coupling}), it is clear that the mass loading $M=\rho_p\langle\phi_p\rangle/\rho_f$ determines the strength of the coupling. Thus, if the mass loading is vanishingly small, the particle phase has little effect on the stability of the Rankine vortex. Conversely, if the mass loading is large, the interaction between two phases triggers a significant source or sink of energy that may enhance or attenuate perturbations in the flow \citep{kasbaouiPreferentialConcentrationDriven2015}.

The particle phase is described in the Lagrangian frame. The motion of the $i$-th particle is given by \citep{maxeyEquationMotionSmall1983}
\begin{eqnarray}
\frac{d\mathbf{x}_p^i}{dt}(t)&=&{\mathbf{u}_p^i}(t),\\
\frac{d\mathbf{u}_p^i}{dt}(t)&=&\frac{1}{\rho_p}\nabla\cdot \tau(\mathbf{x}_p^i,t)+ \frac{\mathbf{u}_f(\mathbf{x}_p^i,t)-\mathbf{u}^i_p(t)}{\tau_p},
\end{eqnarray}
where  $\mathbf{x}_p^i$, and $\mathbf{u}_p^i$, are the position and velocity of the ``$i$''-th particle, respectively, $\tau_p=\rho_pd_p^2/(18\mu)$ is the particle response time, and $d_p$ the particle diameter. In this study, gravity is ignored in order to focus on inertial effects. Other interactions, including collisions, are negligible due to the large density ratio and low volume fraction.
Note that in the equations above, the instantaneous particle volume fraction and Eulerian particle velocity field are obtained from Lagrangian quantities using
\begin{eqnarray}
\phi_p(\mathbf{x},t)&=&\sum_{i=1}^N V_p  g\left(\left|\left|\mathbf{x}-\mathbf{x}_p^i\right|\right|\right), \\
\phi_p \mathbf{u}_p(\mathbf{x},t)&=&\sum_{i=1}^N \mathbf{u}_p^i(t) V_p g\left(\left|\left|\mathbf{x}-\mathbf{x}_p^i\right|\right|\right),
\end{eqnarray}
where $V_p=\pi d_p^3/6$ is the particle volume, $g$ represents a Gaussian filter kernel of size $\delta_f=3\Delta x$, where $\Delta x$ is the grid spacing \citep{capecelatroEulerLagrangeStrategy2013}.

This computational framework was recently applied by the authors in a configuration similar to the one considered here. In \citet{shuaiAcceleratedDecayLambOseen2022}, the dynamics of a particle-laden Lamb-Oseen vortex at moderate Stokes numbers are investigated using the Eulerian-Lagrangian methodology presented here. Readers interested in further details about the numerical approach are referred to this study.

\subsection{Illustration of the instability}

To illustrate the unstable dynamics activated by two-way coupling, we consider a Rankine vortex at the circulation Reynolds number $\mathrm{Re}_\Gamma=\Gamma/(2\pi\nu)=1000$, laden with mono-disperse particles having diameter at the Stokes number {\color{revision} $\Sto=\tau_p/\tau_f=0.025$}, average volume fraction $\langle\phi_p\rangle=1.2\,10^{-3}$, density ratio $\rho_p/\rho_f=830$, and mass loading $M=1$. Here, $\Gamma$ corresponds to the vortex circulation, $r_c$ is the initial vortex core radius, $\tau_f= 2\pi r_c^2/\Gamma$ is the characteristic fluid velocity,  $\tau_p=\rho_p d_p^2/(18\mu_f)$ is the particle response time, and $d_p$ is the particle diameter. The particles are seeded randomly within the core region $r\leq r_c$ with velocities matching the fluid velocity at their respective locations. We use the Eulerian-Lagrangian framework described above on uniform Cartesian grid with resolution $ r_c/\Delta x=50$. The simulations performed here are termed ``pseudo-2D'', meaning that the axial direction $z$ is resolved with one grid point and taken periodic with a thickness $\Delta z=3d_p$. This allows the definition of volumetric quantities such as particle volume and volume fraction. Despite considering pseudo-2D simulations, a large number of particles must be tracker, here $N=376612$, due to the large scale ratio $r_c/d_p=1400$. The remaining numerical parameters are as in \citet{shuaiAcceleratedDecayLambOseen2022}.

In addition to two-way coupled simulations, we perform one-way coupled simulations where the momentum exchange term (\ref{eq:coupling}) is purposely set to zero. In this way, comparing one-way and two-way simulations reveals the effect of particle feed-back on the flow dynamics.

\begin{figure}
  \center
  \includegraphics[width=5.2in]{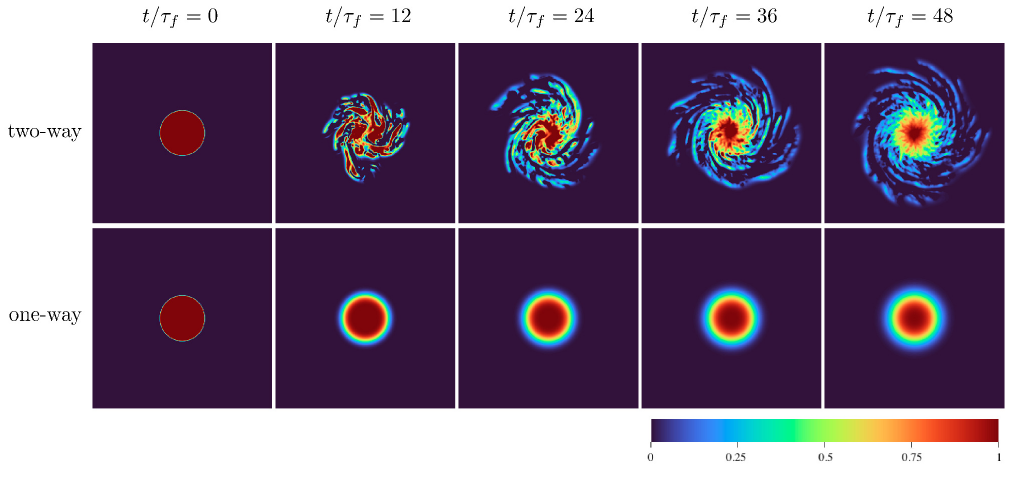}
  \caption{Iso-contours of normalized vorticity ($\omega_z/\omega_c$) with $\Sto=0.025$ and $\mathrm{Re}_{\Gamma}=1000$, for one-way coupling and two-way coupling at five non-dimensional times $t/\tau_f$. \label{fig:evidence_vort}}
\end{figure}

Figure \ref{fig:evidence_vort} shows the evolution of the isocontours of axial vorticity $\omega_z$ normalized by the initial center vorticity $\omega_c=\omega_z(r=0,t=0)$. Successive snapshots are given between $t/\tau_f=0$ and 48. When the particle feedback force is neglected, i.e., in one-way coupling, we observe that the vorticity field remain axisymmetric at all times. Although some viscous diffusion can be observed near the vorticity jump $r\sim r_c$, the vorticity magnitude within the core remains mostly flat as viscous effects are too small to cause significant diffusion within the time frame $0\leq t/\tau_f\leq 48$. Contrary to the case of one-way coupling, we observe significant distortion of the flow field when two-way coupling is accounted for. The vorticity field quickly loses its cylindrical symmetry due to the emergence of azimuthal perturbations. The latter grow into vorticity filaments by $t/\tau_f\sim 12$ that are gradually convected away from the vortex core. By $t/\tau_f\sim 48$, we observe several well-established spiral arms emanating from the vortex core. In this region, vorticity assumes a diffuse profile in contrast to the nearly flat profile seen in one-way coupled simulations.

\begin{figure}
  \center
  \includegraphics[width=5.2in]{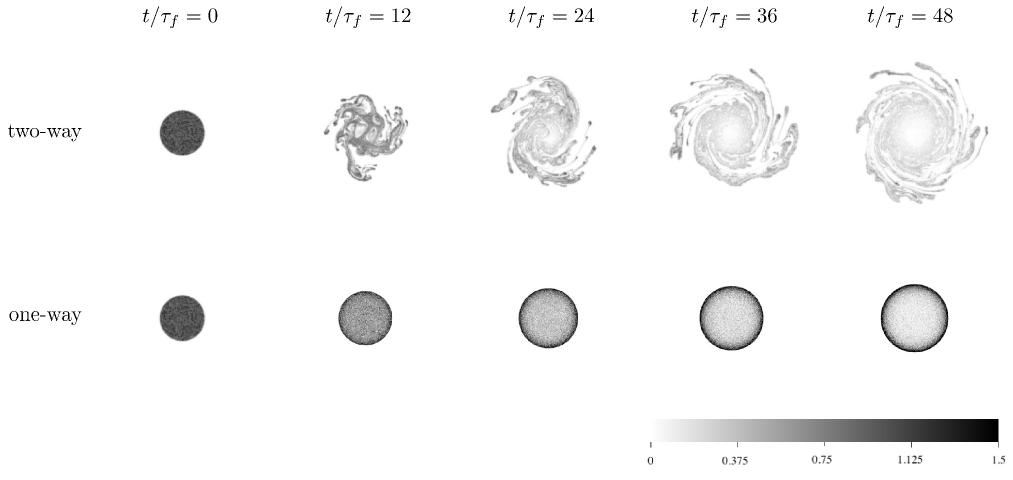}
  \caption{Iso-contours of normalized particle volume fraction ($\phi_p/\langle\phi_p\rangle$) with $\Sto=0.025$ and $\mathrm{Re}_{\Gamma}=1000$, for one-way coupling and two-way coupling at five non-dimensional times $t/\tau_f$.\label{fig:evidence_vfp}}
\end{figure}
    
Particle dispersion is also significantly impacted by the interaction between the two phases. Figure \ref{fig:evidence_vfp} shows the iso-contours of particle volume fraction.
When the particle feedback is neglected, the particles gradually accumulate into a ring-shaped cluster of diameter about $1.5\times r_c$. These clustering dynamics were reported in the earlier work of Druzhinin \citep{druzhininConcentrationWavesFlow1994,druzhininTwowayInteractionTwodimensional1995}, \citet{marshallParticleDispersionTurbulent2005b}, and later by \citet{shuaiAcceleratedDecayLambOseen2022}. The outward particle ejection is due to preferential concentration \citep{squiresPreferentialConcentrationParticles1991} whereby inertial particles are driven out of strongly vortical regions. Like the carrier flow, the particle distribution is axisymmetric only in one-way coupled simulations. In contrast to these prior observations, two-way coupling leads to a loss of symmetry along with fast and wide dispersion of the particles. The two-way coupled simulations in figure \ref{fig:evidence_vfp} reveal the presence of azimuthal perturbations superimposed on the particle patch at $t/\tau_f=12$. The perturbations extend into spiraling particle filaments  reminiscent of density-driven radial Rayleigh-Taylor instability observed by \citet{dixitStabilityVortexRadial2011} and \citet{jolyRayleighTaylorInstability2005}. These perturbations destroy the ring structure observed in one-way coupled simulations.

The simulations shown in figure \ref{fig:evidence_vort} and \ref{fig:evidence_vfp} suggest that semi-dilute inertial particles trigger a modal instability. The latter cannot be attributed to purely hydrodynamic nor purely granular effects since the Rankine vortex is neutrally stable and particle-particle interactions are neglected. Rather, it is an instability that arises from the two-way momentum exchange between the two phases.

The instability observed here represents a distinct mechanism from the one studied by \citet{kasbaouiPreferentialConcentrationDriven2015}. There, the authors show that the interplay between particle settling and preferential concentration activates a non-modal instability in two-way coupled shear flows. It results in the formation of sheets of concentrated particle clusters that rotate to progressively align with the direction of the shear flow. Since particle settling is required for perturbations to grow, the resulting growth rates depend on the gravitational acceleration. In the present work, a modal instability arises regardless of gravitational effects.

\section{Linear stability analysis for weakly inertial particles}
\label{sec:lsa}

In this section, we establish a theoretical grounding to this novel two-phase instability through rigorous linear stability analysis (LSA). The analysis is intended to reveal the  mechanisms that cause {\color{revision} the 2D particle-laden vortex flow to be unstable}, and their functional dependence on the non-dimensional numbers at hand. This study is restricted to the assumption of small, but non-zero particle inertia such that $\Sto<0.1$. 

In the following, we adopt an Eulerian-Eulerian description of the particle-laden flow based on the Two-Fluid model \citep{marbleDynamicsDustyGases1970,druzhininTwowayInteractionTwodimensional1995,jacksonDynamicsFluidizedParticles2000,kasbaouiPreferentialConcentrationDriven2015,kasbaouiRapidDistortionTwoway2019}. In this framework, the conservation equations for the fluid phase and particle phase in the two-fluid model read
\begin{eqnarray}
  \nabla \cdot \bm{u}_f&=&0, \label{eq:TF1}\\
  \rho_f \left(\frac{\partial \bm{u}_f}{\partial t}+\bm{u}_f\cdot\nabla\bm{u}_f\right)&=&-\nabla p +\mu \nabla^2\bm{u}_f + \frac{\rho_p V_p n}{\tau_p}(\bm{u}_p-\bm{u}_f),\label{eq:TF2}\\
  \frac{\partial \rho_p n}{\partial t} +\nabla\cdot \left( \rho_p n\bm{u}_p\right)&=& 0, \label{eq:TF3}\\
  \frac{\partial \rho_p n \bm{u}_p}{\partial t} +\nabla\cdot \left( \rho_p n \bm{u}_p \bm{u}_p\right) &=& - \frac{\rho_p V_p n}{\tau_p}(\bm{u}_p-\bm{u}_f),\label{eq:TF4}
\end{eqnarray}
where $\bm{u}_f$ is the fluid velocity, $\bm{u}_p$ is the particle velocity, $V_p$ is the particle volume and  {\color{revision}$n=\phi_p/V_p$} is the particle number density. Equations (\ref{eq:TF1}) and (\ref{eq:TF2}) express mass and momentum conservation for the fluid phase, respectively. Equations (\ref{eq:TF3}) and (\ref{eq:TF4}) express mass and momentum conservation for the particle phase. In these equations, it is assumed that the particles experience a hydrodynamic force equal to Stokes drag. 
The two-phases are coupled through Stokes drag as seen in equations (\ref{eq:TF2}) and (\ref{eq:TF4}). If the particle feedback force is dropped from the fluid momentum equation (\ref{eq:TF2}), the evolution of the fluid becomes decoupled from that of the particles. The fluid evolves as in single-phase, and the flow may not become unstable as shown in \S \ref{sec:evidence} and \citet{michalkeInviscidInstabilityCertain1967}. The momentum exchange between the two phases is critical for the development of an instability.

Compared to the Eulerian-Lagrangian framework presented in \S \ref{sec:evidence}, the form of the differential equations in the Two-Fluid model is more suitable for theoretical analysis using LSA. Still the two approaches yield qualitatively and quantitatively similar evolution of particle-laden flows in the semi-dilute regime provided that particle inertia is small. Further details on the comparison of the two formulations can be found in  \citep{kasbaouiClusteringEulerEuler2019}. Thus, we restrict the analysis to cases where $\Sto \ll 1$.

Under the assumption of weakly inertial particles, it is possible to further simplify the governing equations by adopting the fast equilibrium approximation (see work by Balachandar and coworkers \citep{ferryEquilibriumExpansionEulerian2002,ferryFastEulerianMethod2001,raniEvaluationEquilibriumEulerian2003} and Maxey \citep{maxeyGravitationalSettlingAerosol1987}). Provided that particle inertia is low enough, it is possible to express the particle velocity field as
\begin{eqnarray}
  \bm{u}_p=\bm{u}_f - \tau_p\left(\frac{\partial \bm{u}_f}{\partial t}+\bm{u}_f\cdot\nabla\bm{u}_f\right)+O(\tau_p^2), \label{eq:small_stokes}
  \end{eqnarray}
where it is seen that the particles deviate from the fluid streamlines by a small inertial correction. Owing to the preferential concentration mechanism, particles suspended in a vortex do not follow the closed-loop streamlines of the carrier flow. Instead, they tend to migrate outward radially at a rate controlled by their Stokes number.  

{\color{revision}Next, we combine equation (\ref{eq:small_stokes}) with the conservation equations (\ref{eq:TF1}), (\ref{eq:TF2}) and (\ref{eq:TF3}).} {\color{revision} Using $r_c$, $\Gamma/(2\pi r_c)$, and the core number density $n_0$ as the reference length scale, velocity scale, and number density,} respectively, the following non-dimensionalized equations are obtained
\begin{eqnarray}
  \nabla \cdot \bm{u}^*&=&0, \label{eq:non_dim_LSA0}\\
  (1+M n^*)\left(\frac{\partial \bm{u}^*}{\partial t^*}+\bm{u}^*\cdot\nabla\bm{u}^*\right)&=&-\nabla p^* +\frac{1}{\mathrm{Re}_{\Gamma}} \nabla^2\bm{u}^*,\label{eq:non_dim_LSA1}\\
  \frac{\partial n^*}{\partial t^*} +\left(\bm{u}^*-\Sto\left( \frac{\partial \bm{u}^*}{\partial t^*}+\bm{u}^*\cdot\nabla\bm{u}^*\right)\right) \cdot \nabla n^*&=&\Sto n^*\nabla\bm{u}^*:\nabla\bm{u}^*, \label{eq:non_dim_LSA2}
\end{eqnarray}
where the stared variables denote non-dimensional quantities, $\mathrm{Re}_{\Gamma}=\Gamma/(2\pi \nu)$ is the circulation Reynolds number, $\Sto=\tau_p r_c^2/\Gamma$ is the circulation Stokes number, and $M=\rho_p V_p n_0/\rho_f$ is the mass loading. {\color{revision} Equation (\ref{eq:non_dim_LSA1}) is derived by injecting the small Stokes expansion (\ref{eq:small_stokes}) in the momentum exchange term in (\ref{eq:TF2}). Likewise, equation (\ref{eq:non_dim_LSA2}) is obtained from (\ref{eq:TF3}) by replacing the particle velocity field with the expansion (\ref{eq:small_stokes}).} For ease of notation, we drop the stars in the rest of the analysis.

In this form, the particle phase is coupled with the fluid through the preferential concentration term appearing in the right hand side of (\ref{eq:non_dim_LSA2}). Since \citet{squiresPreferentialConcentrationParticles1991}, preferential concentration term has been widely studied \citep{squiresPreferentialConcentrationParticles1991,wangSettlingVelocityConcentration1993,batchelorInstabilityStationaryUnbounded1991,druzhininTwowayInteractionTwodimensional1995}. It is understood that this term relates to compressibility effects of the particle phase since $\nabla\cdot \bm{u}_p\simeq -\tau_p\nabla \bm{u}_f:\nabla \bm{u}_f$ for low inertia particles. This term may be further expressed as $\nabla\bm{u}_f:\nabla \bm{u}_f=S^2-R^2$, where $S^2$ and $R^2$ represent the second invariants of the fluid strain rate tensor $S=(\nabla \mathbf{u} +\nabla \mathbf{u}^T)/2$ and rotation tensor $R=(\nabla \mathbf{u} -\nabla \mathbf{u}^T)/2$ \citep{squiresPreferentialConcentrationParticles1991}. When the local strain exceeds the local rotation, this term is positive leading to particle accumulation. Conversely, when this term is negative, the particles are expelled. As a result, inertial particles are progressively depleted from rotational regions, such as vortex cores. The formation of a ring cluster shown in figure \ref{fig:evidence_vfp} is a direct effect of preferential concentration.

The number density term in (\ref{eq:non_dim_LSA1}) suggests that the particles may lead to non-Boussinesq dynamics similar to those observed in stratified fluids. In particular, \citet{dixitVortexinducedInstabilitiesAccelerated2010a} showed that density gradients could lead to a destabilization of 2D vortex. \citet{sippStabilityVortexHeavy2005} showed that density gradients produce two distinct kinds of instability, namely centrifugal instability (CTI) which mainly affects axisymmetric eigenmodes and Rayleigh–Taylor instability (RTI) which mainly affects non-axisymmetric eigenmodes. If particle inertia is negligible, one may ignore the preferential concentration term, and treat the suspension as a fluid with effective density $\rho_{\mathrm{eff}}=\rho_f+\rho_p V_p n$. In such case, the suspension would be susceptible to the instabilities of variable density vortex flows aforementioned. These effects are revisited for the case of non-inertial particles in \S \ref{sec:Sto}, and then extended to the case of weakly inertial particles in \S \ref{sec:effects_PI}.

\subsection{Base state}\label{sec:base_state}

\begin{figure}
  \center
  \begin{subfigure}{0.45\linewidth}
    \includegraphics[width=\linewidth]{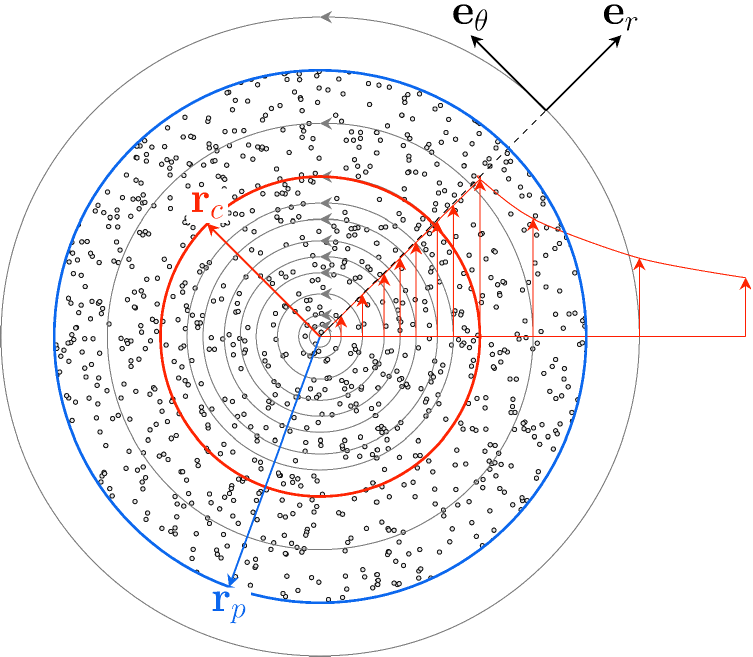}
    \caption{\label{fig:config_a}}
  \end{subfigure}\hfill
  \begin{subfigure}{0.5\linewidth}
    \includegraphics[width=\linewidth]{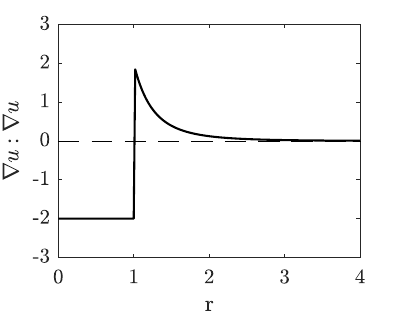}
    \caption{\label{fig:config_c}}
  \end{subfigure}
  \caption{Configuration studied using linear stability analysis: (a) a Rankine vortex with core radius $r_c$ is seeded with particles within the region $r\leq r_p$; (b) the resulting preferential concentration term is negative within the vortex core which causes inertial particles in this region to be ejected outward.\label{fig:configuration_rankine}}
\end{figure}

In the remainder of the manuscript, we perform a linear stability analysis for a base state initially comprising monodisperse particles seeded randomly within a disk of non-dimensional size $r_p$ and a Rankine vortex. A schematic of the configuration is given in figure \ref{fig:config_a}. The non-dimensional base fluid azimuthal velocity field $u^b_{f,\theta}$ and number density fields $n^b$ are given by
\begin{eqnarray}
u^b_{f,\theta}(r,t=0)&=& \left\{
             \begin{array}{lr}
             r, & r\leq1\\
             1/r, & r>1
             \end{array}
            \right.
\label{eq:base_flow}
            \\
n^b(r,t=0)&=& \left\{ 
\begin{array}{l l}
1, & \mbox{if } r\leq r_p\\
0, & \mbox{otherwise.}
\end{array}\right.
\label{eq:base_ND}
\end{eqnarray}

Note that  the task of defining a base state requires careful consideration, when the suspended particles are inertial. Even when the base flow is steady, the particle distribution may be unsteady due to the outward ejection of particles caused by the preferential concentration term in equation (\ref{eq:non_dim_LSA2}). The latter is plotted in figure \ref{fig:config_c} for the Rankine vortex  in equation (\ref{eq:base_flow}). The preferential concentration term is negative within the core ($r<1$) showing that it acts as a sink in this region. Conversely, the preferential concentration term is positive away from the core ($r>1$), corresponding to regions where the particles will accumulate. Thus, the base number density $n^b$ in (\ref{eq:base_ND}) is expected to vary as time progresses.

The time evolution of the base state may be obtained by solving equations (\ref{eq:non_dim_LSA0}), (\ref{eq:non_dim_LSA1}), and (\ref{eq:non_dim_LSA2}). Considering an inviscid and axisymmetric state ($\partial()/\partial \theta = 0$), the equations dictating the evolution of the base state are
\begin{eqnarray} 
    \frac{\partial u_{\theta}^b}{\partial t} = 0, \label{eqn:singlefluid_unsteady_utheta} \\
    \frac{\partial n^b}{\partial t} + \frac{\Sto}{r} \frac{\partial}{\partial r} \left( (u_{\theta}^{b})^2 n^b \right) = 0. \label{eqn:singlefluid_unsteady_n}
\end{eqnarray}
Following \cite{druzhininConcentrationWavesFlow1994}, it is possible to obtain an analytical solution for the unsteady number density by applying the method of characteristics. Denoting the initial particle number density in (\ref{eq:base_ND}) by $n_0^b = n^b(r,0)$, the unsteady number density is written as 
\begin{eqnarray}
n^b(r,t)&=& \left\{ 
\begin{array}{l l}
e ^{-2 \: \Sto \: t} \; n_0^b\left(r e ^{- \Sto \; t}\right), & \mbox{if } r\leq 1\\
\frac{\displaystyle r^2}{\displaystyle\sqrt{r^4 - 4 \Sto \: t}} \; n_0^b\left( (r^4 - 4 \Sto \: t)^{1/4} \right), & \mbox{if } r>1 \,\,\mbox{and}\,\, t\leq t_{\mbox{cr}}\\
\left(rr_0\right)^2\; n_0^b\left( r_0\right), & \mbox{if } r>1 \,\,\mbox{and}\,\, t> t_{\mbox{cr}}
\end{array}\right. \label{eq:unsteady_base_ND}
\end{eqnarray}
where $t_{\mbox{cr}}=(r^4-1)/4\Sto$ and $r_0(r,t)$ is obtained from the equation $4\log r_0+r_0^4=r^4-4\Sto t$. Despite the explicit time-dependence of the base number density field $n^b$, a quasi-steady assumption may be employed for weakly inertial particles.
Under this assumption, the base state is assumed frozen at $t=0$. This assumption is justified provided that perturbations grow on a time scale that is much faster than the characteristic evolution time of the number density field. From  solution (\ref{eq:unsteady_base_ND}), it can be seen that the timescale for the development of number density inhomogeneity is  $\tau_c=\tau_f/\Sto$.  \citet{shuaiAcceleratedDecayLambOseen2022} showed from Eulerian-Lagrangian simulations that the time required to eject most particles from the core of a vortex is $\sim 3\tau_c$. 
This timescale may be significantly larger than the flow inertial timescale $\tau_f$ when $\Sto\ll 1$. For such weakly inertial particles with low inertia, particle clustering is slow compared to inertial effects of the carrier vortex flow. Since we expect the instability to develop on a timescale comparable with the flow inertial timescale, the assumption of quasi-steady number density is justified.

\subsection{Linearized equations}\label{sec:small_st}
We assume that the base state discussed previously is subject to small perturbations $n'$ and $\bm{u}'$ such that the total number density and fluid velocity are $n=n^b+n'$ and $\bm{u}=\bm{u}^b+\bm{u}'$. Taking the inviscid limit ($\mathrm{Re}_{\Gamma} \rightarrow \infty$) and linearizing equations (\ref{eq:non_dim_LSA0}) and (\ref{eq:non_dim_LSA1}), the fluid perturbation is subject to the following equations
\begin{eqnarray}
  \frac{1}{r}\frac{\partial (r u_{r}')}{\partial r}  + \frac{1}{r}\frac{\partial  u_{\theta}'}{\partial \theta}&=&0, \label{eq:perturbation-0}\\
  (1+Mn^b)\left( \frac{\partial u_{r}'}{\partial t} +\frac{ u^b_{\theta}}{r}\frac{\partial u_{r}'}{\partial \theta} - 2 \frac{u^b_{\theta}u'_{\theta}}{r}
  \right)
  - M n' \frac{(u^b_{\theta})^2}{r} &=& -\frac{\partial p'}{\partial r}, \label{eq:perturbation-1}\\
   (1+Mn^b)\left( \frac{\partial u_{\theta}'}{\partial t} +\frac{ u^b_{\theta}}{r}\frac{\partial u_{\theta}'}{\partial \theta} - u'_{r}\frac{\partial u^b_{\theta}}{\partial r}+  \frac{u'_{r}u^b_{\theta}}{r}
  \right)
   &=& -\frac{1}{r}\frac{\partial p'}{\partial \theta}, \label{eq:perturbation-2}
  \end{eqnarray}
where equation (\ref{eq:perturbation-0}) represents mass conservation in cylindrical coordinates, and equations (\ref{eq:perturbation-1}) and (\ref{eq:perturbation-2}) represent the radial and azimuthal fluid momentum conservation respectively. Introducing the base state angular velocity $\Omega^b=u^b_{\theta}/r$ and axial vorticity $\omega^b_z=(\partial (r u^b_{\theta})/\partial r)/r$, equations (\ref{eq:perturbation-1}) and (\ref{eq:perturbation-2}) become
\begin{eqnarray}
 (1+Mn^b)\left( \left(\frac{\partial}{\partial t} + \Omega^b \frac{\partial}{\partial \theta}\right)u'_{r}  - 2\Omega^b u'_{\theta}
  \right)
  - M n' r(\Omega^b)^2 &=& -\frac{\partial p'}{\partial r}, \label{eq:perturbation-3}\\
   (1+Mn^b)\left( \left(\frac{\partial}{\partial t} + \Omega^b \frac{\partial}{\partial \theta}\right)u'_{\theta} +\omega_z^b u'_{r}
  \right)
   &=& -\frac{1}{r}\frac{\partial p'}{\partial \theta}. \label{eq:perturbation-4}
  \end{eqnarray}
The evolution of the perturbation axial vorticity, i.e.,
\begin{eqnarray}
\omega'_z&=&\frac{1}{r}\frac{\partial}{\partial r}(ru'_{\theta})- \frac{1}{r}\frac{\partial u'_{r}}{\partial \theta},
\end{eqnarray}
is obtained by combing equations (\ref{eq:perturbation-3}) and (\ref{eq:perturbation-4}),
\begin{eqnarray}
\left(\frac{\partial}{\partial t} + \Omega^b \frac{\partial}{\partial \theta}\right)\omega'_z + \frac{\partial \omega^b_z}{\partial r} u'_{r}
 &=& \nonumber \\
 && \hspace{-3cm}-\frac{M}{1+Mn^b} \left( (\Omega^b)^2\frac{\partial n'}{\partial \theta}+ \frac{\partial n^b}{\partial r} \left( 
 \left(\frac{\partial}{\partial t} + \Omega^b \frac{\partial}{\partial \theta}\right)u'_{\theta} + \omega^b_z u'_{r}
    \right)
 \right). \label{eq:perturbation-5}
  \end{eqnarray}
The number density perturbation is solution to the following equation,
\begin{eqnarray}
   \frac{\partial n'}{\partial t}+\Omega^b \frac{\partial n' }{\partial \theta}+\frac{\partial n^b}{\partial r} u'_{r} 
   &=& \nonumber\\
   && \hspace{-3.5cm}\Sto\left( \frac{\partial n_b}{\partial r} \left(\frac{\partial u'_{r}}{\partial t}+\Omega^b \frac{\partial u'_{r}}{\partial \theta} \right) -r(\Omega^b)^2 \frac{\partial n'}{\partial r}-2\Omega^b\frac{\partial n^b}{\partial r} u'_{\theta}\right) \nonumber\\
   &&\hspace{-3.5cm}+\Sto n^b\frac{2}{r}\left(\frac{\partial (r\Omega)}{\partial r}\left(\frac{\partial u'_{r}}{\partial \theta}-u'_{\theta} \right)-r\Omega^b \frac{\partial u'_{\theta}}{\partial r}-\Sto\frac{1}{r}\frac{\partial }{\partial r}((r\Omega^b)^2)n' \right). \label{eq:perturbation-6}
\end{eqnarray}

Next, the perturbations $n'$, $u_{\theta}'$, and $u_{r}'$ are decomposed using Fourier series as
\begin{equation} \label{eq:mode}
    \left[u_{r}'(r,\theta,t),u_{\theta}'(r,\theta,t),n'(r,\theta,t)\right] = \left[\hat{u}_r(r,t),\hat{u}_{\theta}(r,t),\hat{n}(r,t)\right] e^{ i m \theta },
\end{equation}
where $m$ represents the mode number. Combining this form of perturbations with equations (\ref{eq:perturbation-5}) and (\ref{eq:perturbation-6}), the set of linear equations {\color{revision} becomes}
\begin{gather}
\mathcal{D} \left( \Bar{n} r \mathcal{D} \left(r \frac{\partial \hat{u}_r}{\partial t}\right) \right) - \Bar{n} m^2 \frac{\partial \hat{u}_r}{\partial t} + i m \Omega^b \left( \mathcal{D} \left( \Bar{n} r \mathcal{D} (r \hat{u}_r) \right) - \Bar{n} m^2 \hat{u}_r \right) \nonumber \\
- i m r \mathcal{D}(\Bar{n} \omega_z^b) \hat{u}_r + r m^2 (\Omega^b)^2 M \hat{n} = 0, \label{eq:coupling_ur_IVP} \\
\left(\frac{\partial n}{\partial t} - \Sto \mathcal{D}(n^{b}) \frac{\partial \hat{u}_r}{\partial t}  \right) + i m \Omega^b \left(n - \Sto \mathcal{D}(n^{b}) \hat{u}_r \right) + \mathcal{D}(n^{b}) \hat{u}_r \nonumber \\
- i \; \Sto \left\lbrace \frac{i}{r} \mathcal{D} (r^2 (\Omega^b)^2 \hat{n}) - \frac{2}{m} \Omega^b \mathcal{D}(n^{b}) \mathcal{D} (r \hat{u}_r) 
\right. \nonumber\\
\left. {} 
- \frac{2 n^b}{m r}  \left[ \mathcal{D} \left( \Omega^b r \mathcal{D} (r \hat{u}_r) \right) - m^2 \mathcal{D}(r \Omega^b) \hat{u}_r \right] \right\rbrace = 0. \label{eq:coupling_n_IVP}
\end{gather}
Here, $\mathcal{D}$ denotes the derivative with respect to $r$ and $\overline{n}=1+Mn^b$. Note that equations (\ref{eq:coupling_ur_IVP}) and (\ref{eq:coupling_n_IVP}) retain the two-way coupling required for an instability. 

Using the assumption of quasi-steady base state discussed above, we perform a modal stability analysis with the base state being frozen in time by considering perturbations that have an exponential dependence on time. Under this consideration, the perturbations can be written as
\begin{equation} \label{eq:norm_mode}
    \left[\hat{u}_r(r,t),\hat{u}_{\theta}(r,t),\hat{n}(r,t)\right] = \left[u_r(r),u_{\theta}(r),n(r)\right] e ^{\sigma t}
\end{equation}
where $\sigma$ is the complex growth rate. Perturbations whose real part of $\sigma$, is positive grow without bound. Thus, the condition for an unstable base state is $\Re(\sigma) >0$. Injecting the forms (\ref{eq:norm_mode}) in equations (\ref{eq:coupling_ur_IVP}) and (\ref{eq:coupling_n_IVP}), we obtain
\begin{gather}
(\sigma + i m \Omega^b) \left[ \mathcal{D} \left( \Bar{n} r \mathcal{D} (r u_r) \right) - \Bar{n} m^2 u_r \right] - i m r \mathcal{D}(\Bar{n} \omega_z^b) u_r + r m^2 (\Omega^b)^2 M n = 0, \label{eq:coupling_ur} \\
(\sigma + i m \Omega^b) \left(n - \Sto \mathcal{D}(n^{b}) u_r \right) + \mathcal{D}(n^{b}) u_r + \Sto \left\lbrace \frac{1}{r} \mathcal{D} (r^2 (\Omega^b)^2 n) + \frac{2 i}{m} \Omega^b \mathcal{D}(n^{b}) \mathcal{D} (r u_r) 
\right. \nonumber\\
\left. {} 
+ \frac{2 i n^b}{m r}  \left[ \mathcal{D} \left( \Omega^b r \mathcal{D} (r u_r) \right) - m^2 \mathcal{D}(r \Omega^b) u_r \right] \right\rbrace = 0. \label{eq:coupling_n}
\end{gather}    
The above pair of equations can be treated as an eigenvalue problem to obtain the stability parameters.

\subsection{Limit of inertia-less particles $(\Sto=0)$ and non-Boussinesq effects} \label{sec:Sto}
As previously discussed in this section, with the absence of particle inertia ($\Sto=0$), the preferential concentration term drops out of the governing equations. Also, the base state in the case of $\Sto=0$ remains constant over time as shown in equations (\ref{eqn:singlefluid_unsteady_utheta}) and (\ref{eqn:singlefluid_unsteady_n}). A simplified linear equations can be obtained as
\begin{gather}
(\sigma + i m \Omega^b) \left[ \mathcal{D} \left( \Bar{n} r \mathcal{D} (r u_r) \right) - \Bar{n} m^2 u_r \right] - i m r \mathcal{D}(\Bar{n} \omega_z^b) u_r + r m^2 (\Omega^b)^2 M n = 0, \label{eq:coupling_ur_St0}\\
n = - \frac{\mathcal{D}(n^b) }{(\sigma + i m \Omega^b)}  u_{r}. \label{eq:coupling_n_St0}
\end{gather} 
The above linearized equations are identical to the equations for a vortex with a radially stratified density written by \cite{dixitStabilityVortexRadial2011}. Without the particle inertia,  the problem reduces to that of a vortex with a density stratification induced by particles. Following the approach of  \cite{dixitStabilityVortexRadial2011}, it is possible solve  equations (\ref{eq:coupling_ur_St0}) and (\ref{eq:coupling_n_St0}) analytically and obtain a dispersion relation for a Rankine vortex. This is done by computing the perturbation velocity and number density fields separately in regions of $r<1$, $1<r<r_p$ and $r>r_p$. The constants subsequently obtained are found using the continuity condition at the jump locations $r=1$ and $r = r_p$. The dispersion relation, thus obtained, is written as
\begin{equation}
    \sigma^3 + a_2 \sigma^2 + a_1 \sigma + a_0 = 0, \label{eqn:Rankine_St0_analytical}
\end{equation}
where
\begin{eqnarray}
    a_2 = i \left( m + 2 m r_p^{-2} - 1 - \Atw \; r_p^{-2m} \right), \\
    a_1 = m r_p^{-2} \left[ 2 - m \left( 2 + r_p^{-2} \right) - \Atw \left( r_p^{-2} - 2 r_p^{-2m} \right) \right], \\
    a_0 = i m (1-m) r_p^{-4} \left[ m + \Atw \left( 1 - r_p^{-2m} \right) \right].
\end{eqnarray}
where $\Atw=M/(2+M)$ is the Atwood number. The latter is a non-dimensional number that commonly arises in density stratified flows. Here, considering the particle-fluid mixture  as a fluid with effective density $\rho_\mathrm{eff}=1+(\rho_p/\rho_f) V_p n=1+M n$, we see that in the core $\rho_\mathrm{eff} (r)= 1+M=\rho_\mathrm{max}$ for $r\leq 1 $, while $\rho_\mathrm{eff}=1=\rho_\mathrm{min}$ away from the core ($r\geq 1$). Thus, the Atwood number is $\Atw=(\rho_\mathrm{max}-\rho_\mathrm{min})/(\rho_\mathrm{max}+\rho_\mathrm{min})$ which measures the relative magnitude of the density jump between the inner and outer regions.

{\color{revision}Figures} \ref{fig:Analytic_contour_o_At_rpbyrc_St0} (a) and (b) show contours of growth rates for wavenumbers $m = 2$ and $4$, respectively, in the $\Atw-r_p$ plane. The white space labelled ``S" represents the region of stability. Thus there exists a critical radius, $r_p^{\textrm{cr}}$, such that instability would only be observed when the density jump is located beyond $r_p^{\textrm{cr}}$. Using the property of the cubic equation (\ref{eqn:Rankine_St0_analytical}) one can obtain a relation for $r_p^{\textrm{cr}}$ as a function of $m$ and $\Atw$, which on further expanding for $\Atw\ll 1$ yields,
\begin{eqnarray}
r_p^{\textrm{cr}} = r^{\textrm{Kelvin}}\left[1-\frac{3}{2}\left\{\frac{\Atw}{4m^2}\left(\frac{m-1}{m}\right)^{m-1}\right\}^{1/3}\right]+\mathcal{O}(\Atw^{2/3})
\end{eqnarray}
where $r^{\textrm{Kelvin}}=\sqrt{m/(m-1)}$ is the critical radius corresponding to the discrete Kelvin mode \citep{roy2014linearized}. As is evident from the complete numerical solution and the above {\color{revision} small-$\Atw$} asymptotic expression, the region of stability shrinks as the wavenumber increases, suggesting that higher modes are more unstable. In addition, higher values of growth rates are achieved with increasing values of Atwood number and with the value of $r_p$ closer to one.  This suggests that the configuration at $r_p=1$ with infinitely heavy core ($M\rightarrow \infty$) is the most unstable configuration.

To further study the influence of the Atwood number on the critical wavenumber, we investigate the specific case of $r_p = 1$, for which the dispersion relation reduces to
\begin{eqnarray}
    \sigma &=& \frac{i}{2}(1 + \Atw) - i m \pm \frac{i}{2}\sqrt{(1 + \Atw)^2 - 4 m \Atw}.
        \label{eqn:omega}
\end{eqnarray}
For $r_p = 1$, the third root of equation (\ref{eqn:Rankine_St0_analytical}) corresponds to a mode rotating with core angular frequency ($\sigma=-im$.) The expression (\ref{eqn:omega}) indicates the critical wavenumber ($m_\mathrm{cr}$) for unstable system is
\begin{gather}
    m_\mathrm{cr} = \frac{(1 + \Atw)^2}{4 \Atw} = \frac{(1+M)^2}{(2+M)M}.
    \label{eqn:critical wave number}
\end{gather}
The above expression shows that increasing the Atwood number, or equivalently the mass loading, promotes the instability by making lower wavenumber modes unstable. In the limit case where $M\rightarrow \infty$, corresponding to $\Atw\rightarrow 1$, the critical wavenumber becomes $m_\mathrm{cr}\rightarrow 1$ showing that all modes are unstable. Conversely, when $M\rightarrow 0$, none of the modes are unstable since $m_\mathrm{cr}\rightarrow \infty$. Thus, we recover the known behavior for an unladen Rankine vortex ($M\rightarrow 0$).

From expression (\ref{eqn:omega}), the growth rate of unstable modes ($m\geq m_\mathrm{cr}$) is given by
\begin{equation}
    \Re(\sigma)=\frac{1}{2}\sqrt{4m \Atw-(1+\Atw)^2}\simeq \sqrt{m \Atw} \quad \text{for } m\gg 1.
\end{equation}
From this relationship, we draw two conclusions. First, a naturally developing instability, i.e., unforced, will tend to emerge at high wavenumbers since $\Re(\sigma)$ increases with $m$. Viscous effects are expected to curb $\Re(\sigma)$ beyond a certain mode number $m$. Because viscous effects are not accounted for in the present formulation, it is not possible to estimate a priori which mode would emerge naturally. The second conclusion is that the growth rate increases with mass loading. The maximum growth rate obtained when $M\rightarrow\infty$ ($\Atw\rightarrow 1$) is $\Re(\sigma)_\mathrm{max}=\sqrt{m}$.

\subsection{Effects of particle inertia} \label{sec:effects_PI}
\begin{figure}
  \centering
  \begin{subfigure}{0.45\textwidth}
    \centering
    \includegraphics[width=1.05\textwidth]{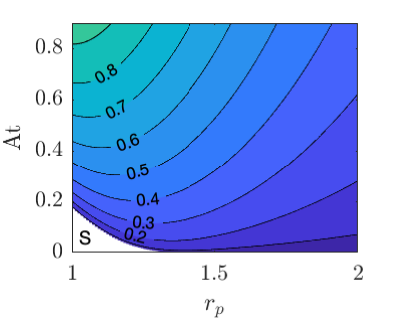}
    \caption{$m = 2$}
  \end{subfigure} 
  \begin{subfigure}{0.45\textwidth}
    \centering
    \includegraphics[width=1.05\textwidth]{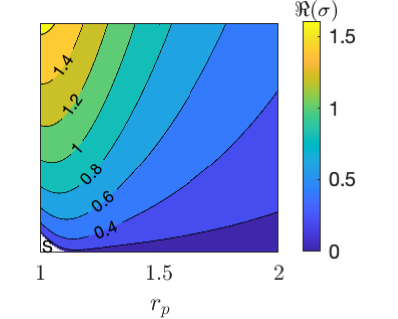}
    \caption{$m = 4$}
  \end{subfigure} 
  \caption{Contours of the analytically obtained growth rate (equation (\ref{eqn:Rankine_St0_analytical})) in the $\Atw-r_p$ plane for a Rankine vortex with $\Sto = 0$.}
  \label{fig:Analytic_contour_o_At_rpbyrc_St0}
\end{figure}

Next, we consider the effects of non-zero particle inertia  ($\Sto \neq 0$) on the stability of the particle-laden Rankine vortex. For this, we first perform a modal stability analysis by solving equations (\ref{eq:coupling_ur}) and (\ref{eq:coupling_n}) with a frozen base state. 

\begin{figure}
  \centering
  \begin{subfigure}{0.49\textwidth}
    \centering
    \includegraphics[width=2.65in]{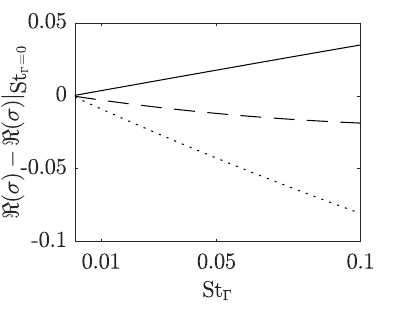}
    \caption{}
  \end{subfigure}
  \begin{subfigure}{0.49\textwidth}
    \centering
    \includegraphics[width=2.65in]{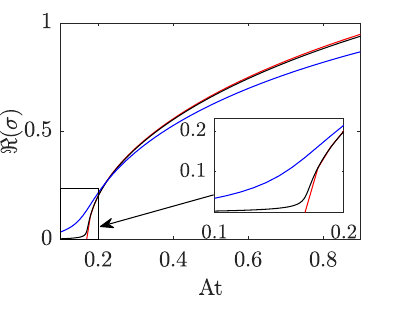}
    \caption{}
  \end{subfigure} 
\caption{Growth rates $\Re(\sigma)$ for a Rankine vortex with $r_p = 1$ and $m = 2$ obtained analytically (equation (\ref{eqn:Rankine_r1-r2_analytical})), (a)  $\Re(\sigma) - \Re(\sigma)|_{\Sto = 0}$ plotted over a range of Stokes numbers for $\Atw = 0.1$ (---), $\Atw = 0.3$ (- - -), $\Atw = 0.9$ ($\cdots$); (b) $\Re(\sigma)$ plotted over a range of Atwood numbers for $\Sto = 0$ (red), $\Sto = 0.01$ (black) and $\Sto = 0.1$ (blue).}
\label{fig:Analytic_vs_num_St001_m2_ci_At}
\end{figure}

For the case where the Rankine vortex and particle core have equal radii, i.e.,  $r_p=1$, a dispersion relation can be obtained analytically:
\begin{equation}
    \sigma^2 - i \left( 1 + \Atw - 2 m + i m \Atw \Sto \right) \sigma - m (m - 1 - i (m-2) \Atw \Sto) = 0, \label{eqn:Rankine_r1-r2_analytical}
\end{equation}
which yields
\begin{equation}
    \sigma = \frac{i}{2} (1+\Atw + i \Atw \Sto m) - i m \pm \frac{i}{2}\sqrt{ (1+\Atw + i \Atw \Sto m)^2 -4 m (\Atw+2i\Atw\Sto)} \label{eq:omega_w_St}
\end{equation}
Note that expression (\ref{eq:omega_w_St}) collapses on (\ref{eqn:omega}) when $\Sto\rightarrow 0$. The corresponding expressions for the eigenfunctions of a mode $m$ are
\begin{eqnarray}
u_r&=& \left\{ 
\begin{array}{l l}
r^{(m-1)}, & \mbox{if }r \leq 1\\
r^{-(m+1)}, & \mbox{otherwise}
\end{array}\right.
\label{eqn:perturbation_initial_velocity}
\\
 n &=& \left( - \Sto + \frac{1 + 2 i \; \Sto}{\sigma + i m} \right) \delta (r-1).
 \label{eqn:perturbation_initial_number}
\end{eqnarray}
Here, $\delta (r-1)$ denotes the Dirac delta function. These functions correspond to perturbations that are concentrated at the number density and vorticity jumps $r=r_p=1$.  To further study the stabilizing/destabilizing effect of particle inertia, we expand the expression for growth rate (equation (\ref{eq:omega_w_St})) in a series of $\Sto$ as
\begin{eqnarray}
\sigma = \sigma\big\vert_{\Sto=0} -\frac{m \Sto\Atw}{2}\pm \left\{\frac{m \Atw(3-\Atw)}{2\sqrt{\Delta}}\Sto+\frac{i m^2 \Atw^2 ((m-2)\Atw+2)}{\Delta^{3/2}}\Sto^2\right\}+\mathcal{O}(\Sto^3).\nonumber \\
\end{eqnarray}
where $\Delta=(1 + \Atw)^2 - 4 m \Atw$. For $\Sto=0$, $\Delta<0$ provides the condition for instability for a fixed $m$, that is, a mode $m$ is stable if  $\Atw<\Atw_{\textrm{cr}}=2m-1-2\sqrt{m(m-1)}$. For $\Sto\neq0$, particle inertia can play contrasting roles. For $\Atw<\Atw_{\textrm{cr}}$, the dispersed phase destabilizes the vortex
\begin{eqnarray}
 \Re(\sigma)=\frac{m\Atw\Sto}{2}\left\{\frac{3-\Atw}{\sqrt{\Delta}}-1\right\}+\mathcal{O}(\Sto^3), \label{eq:atp_smallst}
\end{eqnarray}
while for $\Atw>\Atw_{\textrm{cr}}$, the dispersed phase stabilizes the vortex
\begin{eqnarray}
 \Re(\sigma)=\frac{1}{2}\sqrt{|\Delta|} - \frac{m\Sto\Atw}{2}+\mathcal{O}(\Sto^2).\label{eq:atm_smallst}
\end{eqnarray}
Figure \ref{fig:Analytic_vs_num_St001_m2_ci_At} (a) shows the variation of growth rate plotted over a range of Stokes numbers for wavenumber $m = 2$. With the smaller value of Atwood number $\Atw = 0.1$, the growth rates decrease with increasing Stokes numbers. However, the trends get reversed for higher Atwood numbers, with the growth rates increasing with increasing Stokes numbers, confirming the predictions from the asymptotic calculations (equations (\ref{eq:atp_smallst}) and (\ref{eq:atm_smallst})). Figure \ref{fig:Analytic_vs_num_St001_m2_ci_At} (b) plots the growth rates over a range of Atwood numbers for three values of Stokes numbers. It is found that the growth rate is always positive as long as the Atwood number $\Atw$ and Stokes number $\Sto$ are non-zero. This indicates that destabilization of the vortex by the particles is guaranteed as long as the particle inertia is non-zero. 

\begin{figure}
  \centering
  \begin{subfigure}{0.49\textwidth}
    \centering
    \includegraphics[width=2.65in]{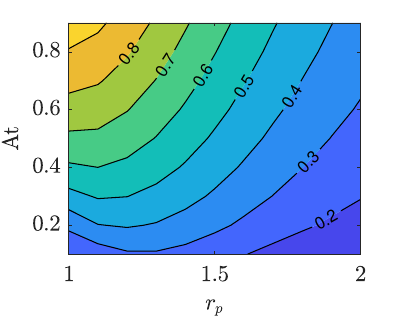}
    \caption{$\Sto = 0.01$}
  \end{subfigure}
  \begin{subfigure}{0.49\textwidth}
    \centering
    \includegraphics[width=2.65in]{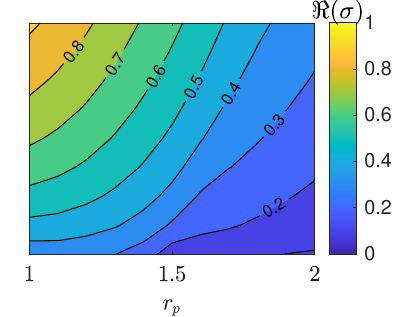}
    \caption{$\Sto = 0.1$}
  \end{subfigure} 
  \caption{Growth rates $\Re(\sigma)$ of mode $m=2$ in the $\Atw$-$r_p$ plane obtained numerically using a smoothed Rankine profile.}
  \label{fig:Contour_tanh_At_beta}
\end{figure}

For cases where $r_p> 1$, we solve equations (\ref{eq:coupling_ur}) and (\ref{eq:coupling_n}) numerically using Chebyshev spectral collocation method \citep{trefethen2000spectral}. To handle the discontinuities, we smooth the base state using  hyperbolic tangents of thickness $\Delta=0.01$. Figure \ref{fig:Contour_tanh_At_beta} (a) and (b) show the contours of growth rate in $\Atw - r_p$ plane for Stokes numbers $0.01$ and $0.1$ respectively. Unlike the case of $\Sto = 0$ (see figure \ref{fig:Analytic_contour_o_At_rpbyrc_St0}), there is no region of stability indicating that mode $m=2$ is always unstable when the particles are inertial. Further, the particle-laden vortex is unstable even for  values of $r_p$ larger than 1. Whereas the results for non-inertial particles are similar to those obtained by \cite{dixitStabilityVortexRadial2011}, inclusion of particle inertia breaks the equivalence with a radially startified Rankine vortex. For the case $\Sto=0.01$, the optimal value of $r_p$ leading to the largest growth rate is slightly in excess of 1 for low values of $\Atw$, and approaches $r_p=1$ as $\Atw$ increases. For the larger inertia case $\Sto=0.1$, $r_p=1$ leads to the largest growth rates for all values of $\Atw$.

\section{Comparison of Linear Stability Analysis with Euler-Lagrange simulations} 
\label{sec:lsa_EL_comp}

In order to validate the linear stability analysis, we compare the analytical results with those from  Eulerian-Lagrangian simulations of the two-way coupled system. To make the comparison tractable and accessible, the case $r_p=1$ is used as the benchmark since the growth rates can be computed analytically from equation (\ref{eqn:Rankine_r1-r2_analytical}). 

Table \ref{tab:mode-non-dim} lists a summary of non-dimensional parameters for 8 cases presently considered. In all these simulations, the Stokes number, circulation Reynolds number $\mathrm{Re}_{\Gamma}$ and ratio $r_p/r_c$ are fixed at $0.002$, $5000$ and $1$, respectively. The Stokes number is sufficiently small such that the initial state can be considered quasi-steady and comparisons with LSA are permissible. In dimensional terms, the carrier fluid has density $\rho_f=1.2\;\mathrm{kg/m^3}$ and viscosity $\mu_f = 1.8 \times 10^{-5}\;\mathrm{kg\cdot m^{-1}\cdot s^{-1}}$. The initial vortex core is $r_c=0.385\ \mathrm{m}$ and circulation is $\Gamma=9.42 \times 10^{-2}\;\mathrm{m^{2}\cdot s^{-1}}$. Particles are seeded with density $\rho_p = 1000\;\mathrm{kg/m^3}$ and average volume fraction $\langle\phi_p\rangle=1.2\times10^{-3}$.  In all cases chosen, the ratio $r_c/d_p$ is large (of the order $10^4$), such that fluctuations due to the discrete particle forcing fall well below the viscous dissipation scale. In this way, the coupling between the fluid and particles occurs primarily through collective particle dynamics at scales comparable with $r_c$, rather than discrete effects at scales comparable with the inter-particle distance. 

\begin{figure}
  \centering
  \begin{subfigure}{0.49\linewidth}
    \includegraphics[width=2.65in]{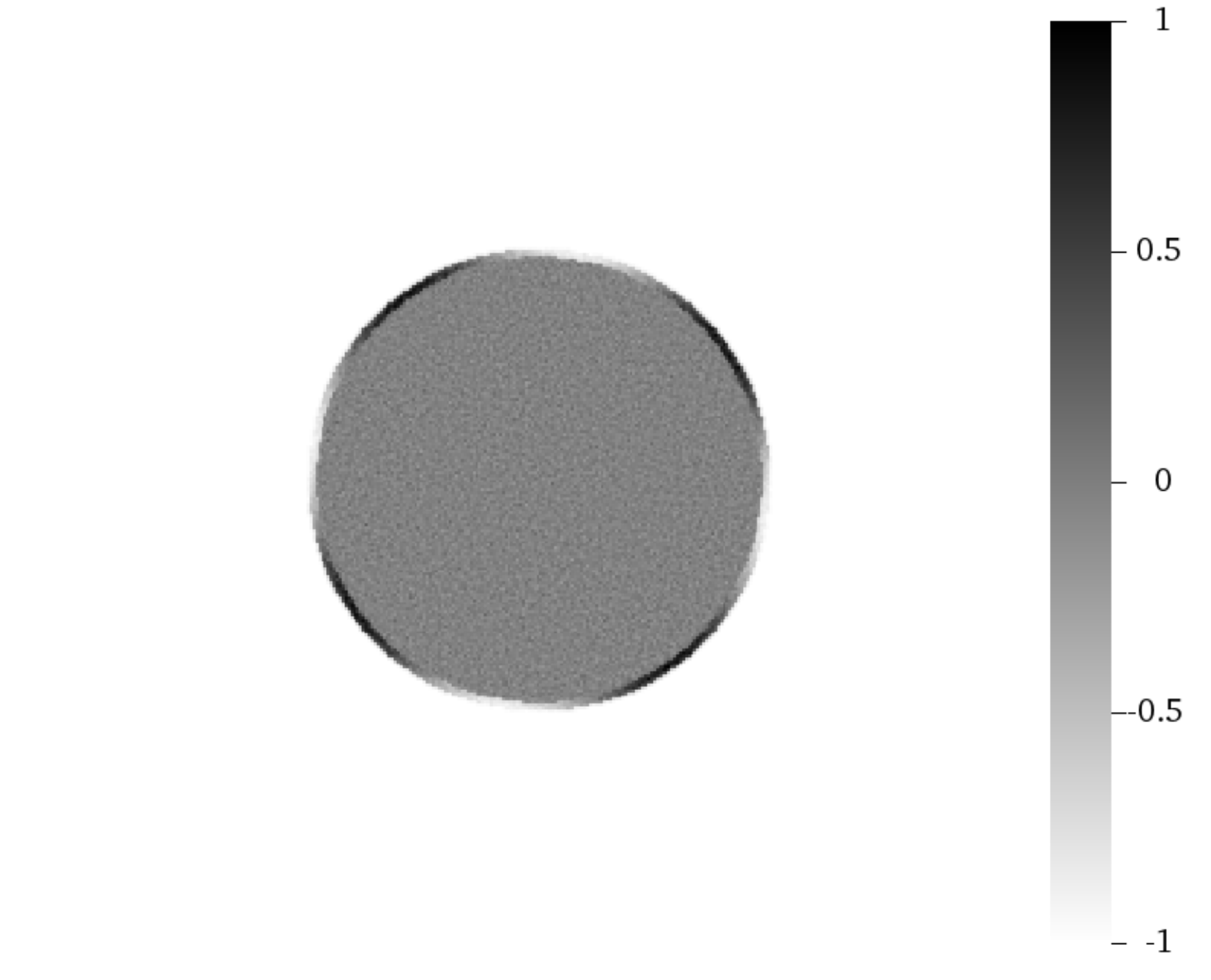}
    \caption{\label{fig:ini_a}}
  \end{subfigure}
  \begin{subfigure}{0.49\linewidth}
    \includegraphics[width=2.65in]{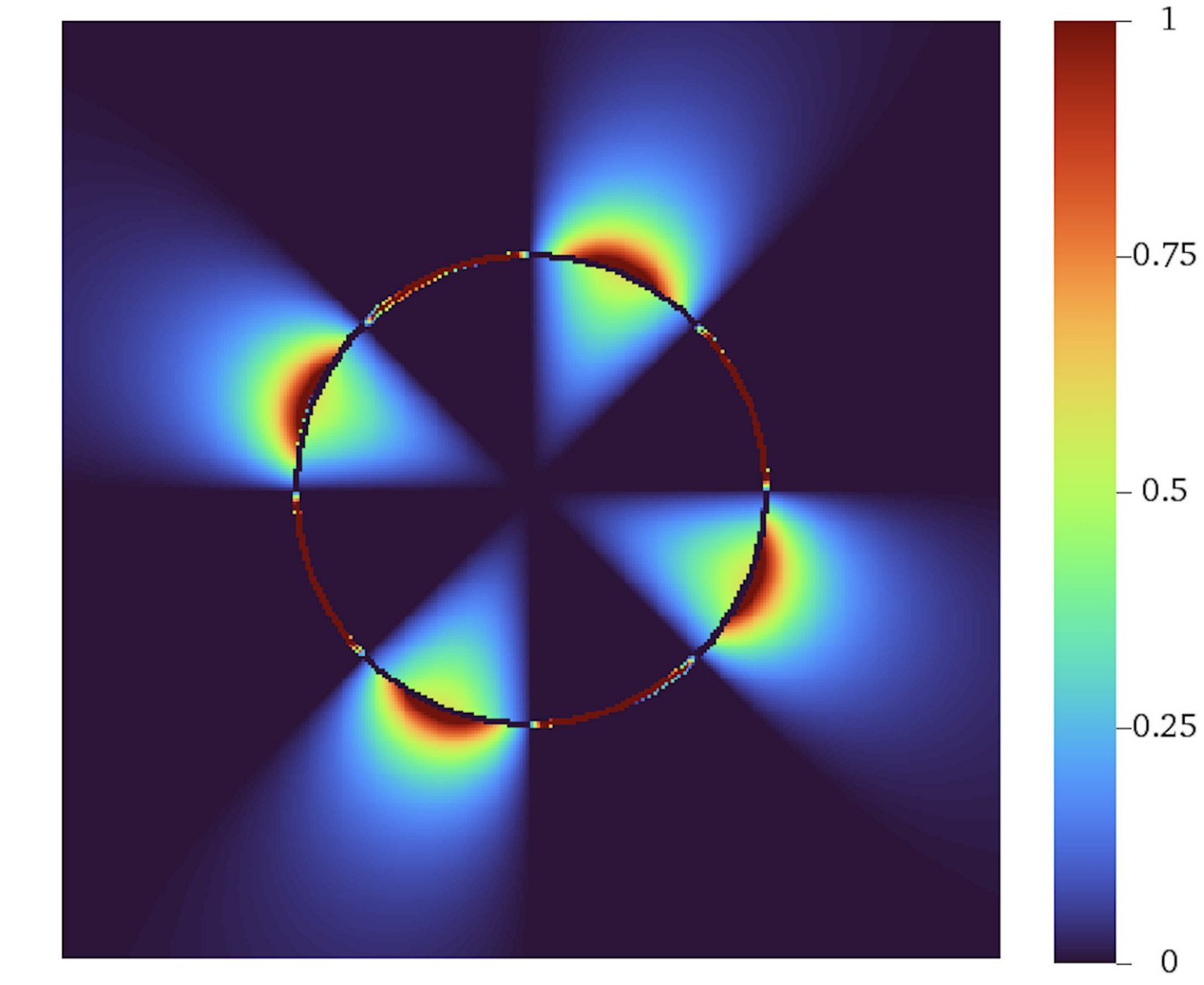}
    \caption{\label{fig:ini_b}}
  \end{subfigure}
  \caption{Example of initial perturbation in the Eulerian-Lagrangian simulations. Normalized number density (a) and axial vorticity (b) perturbations based on eigenmodes (\ref{eqn:perturbation_initial_velocity}) and (\ref{eqn:perturbation_initial_number}) with mode number $m=4$. \label{fig:ini}}
\end{figure}

The initial conditions of the numerical simulations correspond to a superposition of the base Rankine vortex and the perturbation eigenmodes (\ref{eqn:perturbation_initial_velocity}) and (\ref{eqn:perturbation_initial_number}). In order to capture the linear regime described by the linear stability analysis, the perturbations are initialized with a small amplitude $\epsilon=0.03$. The Dirac delta function that appears in the eigenfunctions (\ref{eqn:perturbation_initial_velocity}) and (\ref{eqn:perturbation_initial_number}) is discretized on the Cartesian grid according to the radial distance $l_i$ (i=1,2,3,4) between the vortex core and the four vertices of each cell. Cells that are intersected by the discontinuity ($\max (l_i)>1$ and $\min (l_i)<1$) have $\delta(r-1)=1/(\Delta x\Delta y)$, whereas  $\delta(r-1)=0$ everywhere else.  The Lagrangian particles are generated randomly within each cell based on the number density computed with the aforementioned discrete perturbation. The particle velocities are set to match the fluid velocity at their locations. Figures \ref {fig:ini_a} and  \ref {fig:ini_b} show number density and axial vorticity isocontours of  a representative perturbation with mode $m=4$ and disturbance amplitude {\color{revision} $\epsilon=0.03$} .

\begin{table}
  \begin{tabularx}{\linewidth}{>{\hsize=2.2\hsize}X>{\hsize=0.8\hsize}XXXXXXX}
  Case &     $m$       &  $\mathrm{Re}_\Gamma$ & $\mathrm{At}$ &  $M$ & $\langle\phi_p\rangle$ & $r_c/d_p$  \\[1ex]
  A     & 2                  & $5000$                              & 1/3                       & 1                       & $1.2\;   10^{-3}$                   & 10760             \\
  B     & 3                  & $5000$                              & 1/3                       & 1                       & $1.2\;   10^{-3}$                   & 10760            \\
  C     & 4                  & $5000$                              & 1/3                       & 1                       & $1.2\;   10^{-3}$                   & 10760             \\
  D     & 5                  & $5000$                              & 1/3                       & 1                       & $1.2\;   10^{-3}$                   & 10760             \\
  E     & 6                  & $5000$                              & 1/3                       & 1                       & $1.2\;   10^{-3}$                   & 10760             \\
  F     & 3                  & $5000$                              & 1/6                       & 2/5                     & $1.2\;   10^{-3}$                   & 10760              \\
  G     & 3                  & $5000$                              & 1/5                       & 1/2                     & $1.2\;   10^{-3}$                   & 10760             \\
  H     & 3                  & $5000$                              & 1/4                       & 2/3                     & $1.2\;   10^{-3}$                   & 10760
  \end{tabularx}
  \caption{Non-dimensional parameters considered in simulations. \label{tab:mode-non-dim}}
\end{table}

\begin{figure}
  \centering
  \includegraphics[width=5.0in]{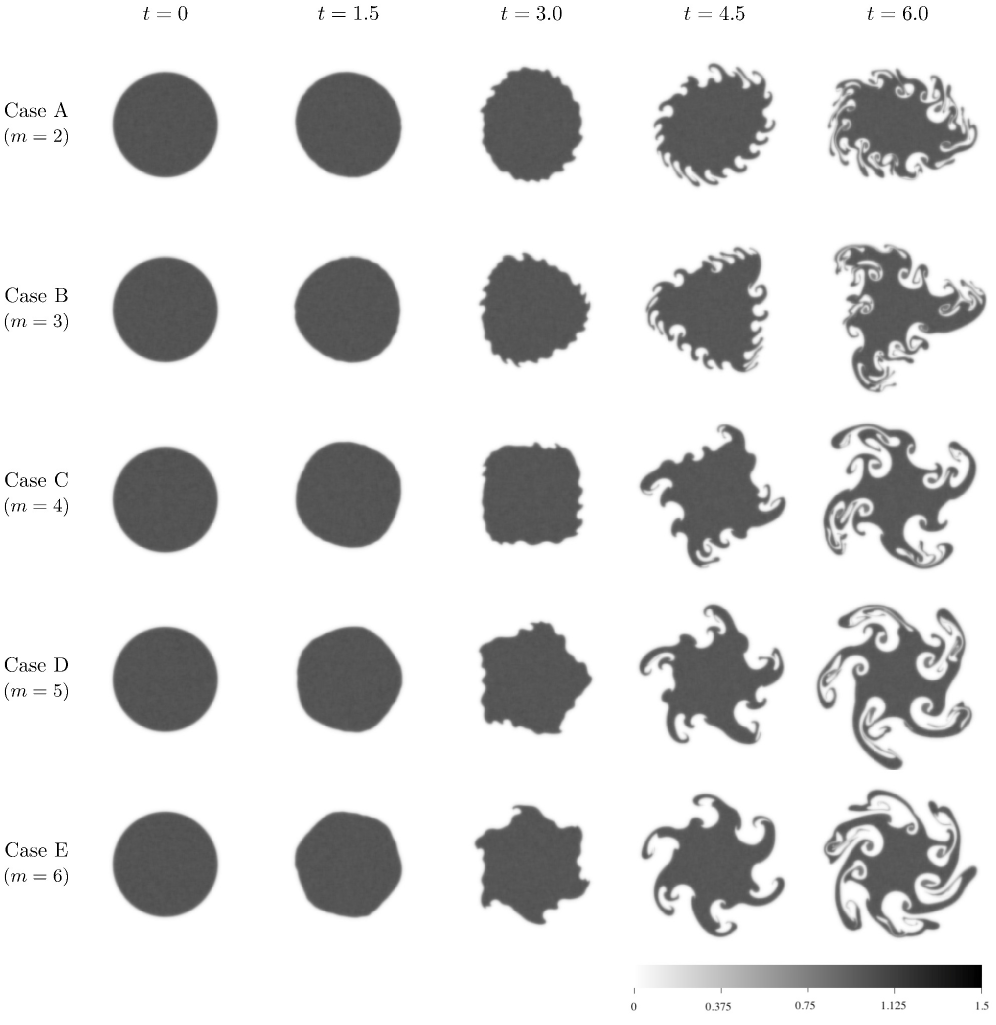}
  \caption{\color{revision2}Iso-contours of normalized particle volume fraction ($\phi_p/\langle\phi_p\rangle$) for cases A, B, C, D, E (see table \ref{tab:mode-non-dim}) at five instant time. As time progresses, the inertial particles are expelled out of the vortex cores, forming cluster arms of which the number equals to the mode number. The small perturbations continue growing and ultimately destroy the structure. \label{fig:vol_frac_m246}}
\end{figure}

\begin{figure}
  \centering
  \begin{subfigure}{0.49\linewidth}
    \includegraphics[width=2.65in]{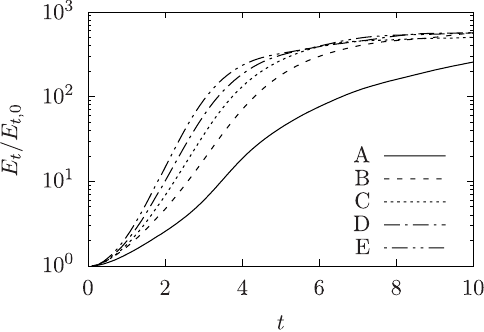}
    \caption{\label{fig:Egrowtha}}
  \end{subfigure}
    \begin{subfigure}{0.49\linewidth}
    \includegraphics[width=2.65in]{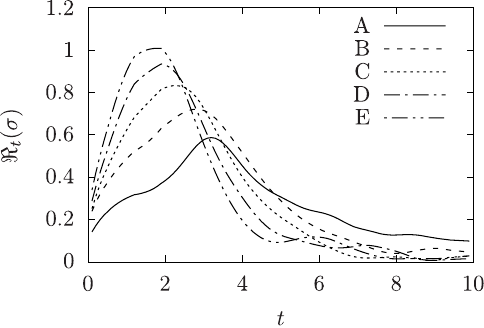}
  \caption{\label{fig:Egrowthb}}
  \end{subfigure}
  \caption{\color{revision2}Time evolution of the perturbation's (a) normalized kinetic energy and (b) growth rate for cases A, B, C, D, and E, corresponding to a Rankine vortex excited with eigenmodes $m=$ 2, 3, 4, 5, and 6, respectively (see table \ref{tab:mode-non-dim}). \label{fig:Egrowth}}
\end{figure}

\begin{figure}
  \centering
  \begin{subfigure}{0.49\linewidth}
    \includegraphics[width=2.65in]{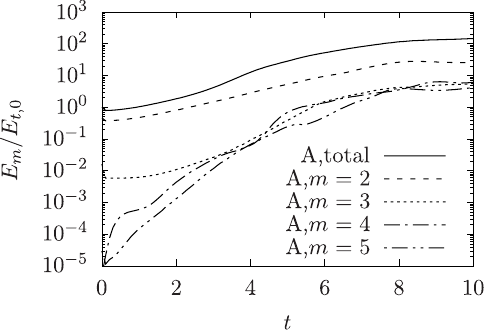}
    \caption{\label{fig:mEgrowtha}}
  \end{subfigure}
  \begin{subfigure}{0.49\linewidth}
    \includegraphics[width=2.65in]{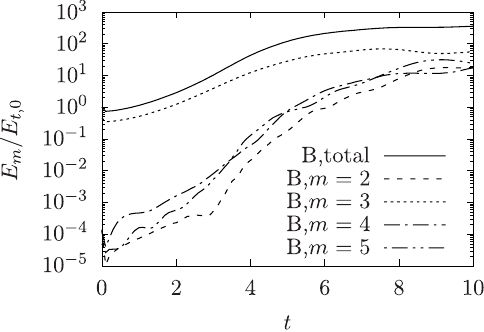}
  \caption{\label{fig:mEgrowthb}}
  \end{subfigure}
  \begin{subfigure}{0.49\linewidth}
    \includegraphics[width=2.65in]{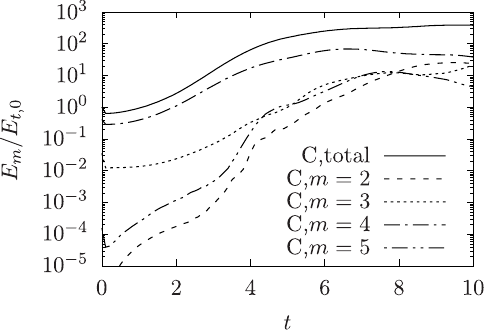}
  \caption{\label{fig:mEgrowthb}}
  \end{subfigure}
      \begin{subfigure}{0.49\linewidth}
    \includegraphics[width=2.65in]{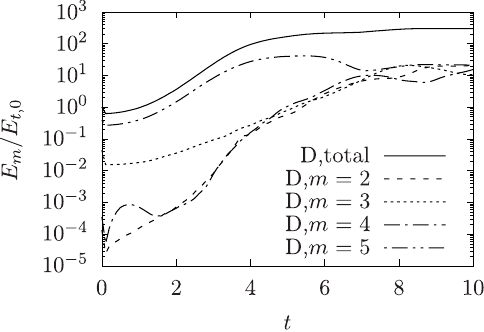}
  \caption{\label{fig:mEgrowthb}}
  \end{subfigure}
  \caption{\color{revision2}
  Evolution of the kinetic energy of the total perturbation and the kinetic energy associated with  azimuthal modes $m=2$, 3, 4, and 5 in: (a) case A (seeded with $m=2$), (b) case B (seeded with $m=3$), (c) case C (seeded with $m=4$), and (d) case D (seeded with $m=5$). The seeded mode dominates during the early evolution $t\lesssim 4$.
  \label{fig:mEgrowth}}
\end{figure}

\begin{figure}
  \centering
  \begin{subfigure}{0.49\linewidth}
    \includegraphics[width=2.65in]{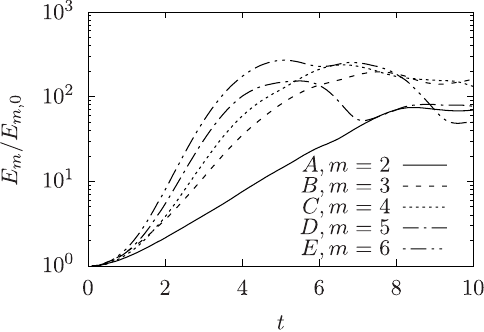}
    \caption{\label{fig:mEgrowth1}}
  \end{subfigure}
    \begin{subfigure}{0.49\linewidth}
    \includegraphics[width=2.65in]{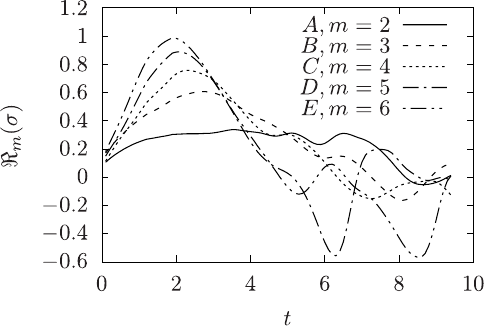}
  \caption{\label{fig:mEgrowth2}}
  \end{subfigure}
  \caption{\color{revision2}Time evolution of seeded mode's (a) normalized kinetic energy and (b) growth rate from cases A, B, C, D, and E.\label{fig:mEgrowth_2}}
\end{figure}

\begin{figure}
  \centering
  \begin{subfigure}{0.49\linewidth}
    \includegraphics[width=2.65in]{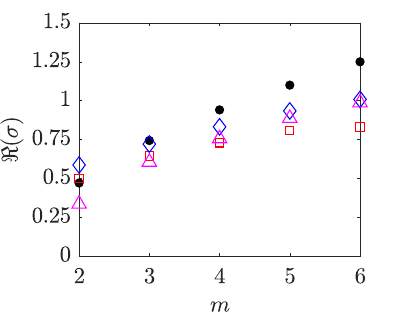}
    \caption{\label{fig:compare_a}}
  \end{subfigure}
  \begin{subfigure}{0.49\linewidth}
    \includegraphics[width=2.65in]{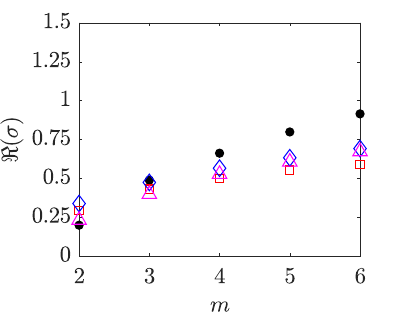}
  \caption{\label{fig:compare_b}}
  \end{subfigure}
    \begin{subfigure}{0.49\linewidth}
    \includegraphics[width=2.65in]{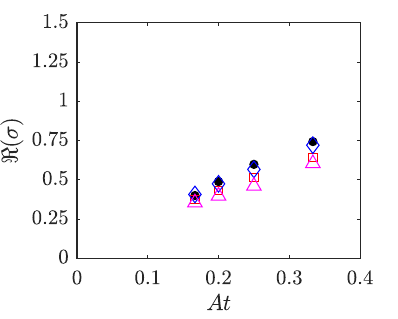}
  \caption{\label{fig:compare_c}}
  \end{subfigure}
  \caption{\color{revision2}Comparison of instability growth rates between linear stability analysis (LSA) and Eulerian-Lagrangian (EL) simulations at $\Sto=0.002$ and $\mathrm{Re}_{\Gamma}=5000$. (a) Growth rates for mode numbers $m=$2 -- 6 at $\mathrm{At}=1/3$ ($M=1$). (b) Growth rates for mode numbers $m=$2 -- 6 at $\Atw=1/5$ ($M=1/2$). (c) Growth rates for mode $m=3$ at various Atwood numbers.
  Symbols: \FilledSmallCircle, LSA; \Diamondshape, EL-peak growth rate of the total perturbation energy; \SmallSquare, EL-average growth rate of the total perturbation energy;  \SmallTriangleUp, EL-peak growth rate of the seeded mode.
  \label{fig:compare}}
\end{figure}

{\color{revision}Figure \ref{fig:vol_frac_m246}} shows the time evolution of the particle volume fraction between $t=0$ and $t=6$ for modes $m=2 - 6$. {\color{revision} All time values in this section are non-dimensionalized with vortex response time $\tau_f$.  For $t\lesssim 3.0$, the perturbed modes grow linearly resulting in the deformation of the interface at the number density jump. The latter, which  is initially circular, develops azimuthal perturbations depending on which mode has been triggered initially. Around $t\simeq 3.0$, we notice the emergence of fast growing high wavenumber modes $m\sim 24-26$ that quickly overtake the initially seeded low wavenumber modes by $t=4.5$. These high wavenumber perturbations are triggered by the discrete Lagrangian particle forcing and errors associated with the discretization of the discontinuous eigenfunctions (\ref{eqn:perturbation_initial_velocity}) and (\ref{eqn:perturbation_initial_number}). 
As shown by the linear stability analysis, the perturbation growth rate increases with wavenumber $m$, such that it may be anticipated that such modes will eventually dominate. Thus, whereas the early-stages are controlled by the seeded low-wavenumber modes, the later-stages are dominated by the high-wavenumber perturbations. By $t=6$, these modes evolve into long spiraling arms of inertial particles that expand radially outward, which indicates strong non-linearity. Regions of unladen fluid in between the particle arms are sucked inward forming mushroom-like structures. The fact that the final stages seen in figure \ref{fig:vol_frac_m246} show similar number density patterns irrespective of the seeded mode, suggests that the mode that would emerge naturally is a high wavenumber mode that results from the balance between two-way coupling and viscous dissipation.}

These dynamics may be related to the misalignment between the particle density interface and centripetal acceleration field \citep{jolyRayleighTaylorInstability2005}, {\color{revision}which is similar to the mechanism of wave interaction between density interface and vorticity interface \citep{dixitStabilityVortexRadial2011}}. The baroclinic torque caused by the misalignment concentrates vorticity on the spiral arms and destroys it in between, resulting in radial filaments.

In order to determine the {\color{revision}total} perturbation growth rates from the present simulations, we compute the perturbation kinetic energy,
 \begin{equation}
    E=\iint\frac{1}{2} ( \bm{u}-\bm{u}_f^b)\cdot ( \bm{u}-\bm{u}_f^b) dS
\end{equation}
where $\bm{u}_f^b$ is the base state velocity corresponding to a Rankine vortex. {\color{revision}Figure \ref{fig:Egrowtha}} shows the evolution of total perturbation energy in {\color{revision}logarithmic} scale for simulations with perturbed modes $m=2$ --6.
{\color{revision}The perturbation energy is seen to increase with increasing mode number $m$, reflecting the larger growth rate for higher mode numbers.}

{\color{revision}The instability growth rate is obtained from Eulerian-Lagrangian simulations using the relation
\begin{equation}
    \Re(\sigma) = \frac{1}{2 E} \frac{d E}{d t},\label{eq:growth_rate}
\end{equation}
Figure \ref{fig:Egrowthb} shows the time evolution of the growth rate given by (\ref{eq:growth_rate}). Unlike the linear stability analysis which predicts a constant value, the instability growth rate varies with time in Euler-Lagrange simulations. Figure \ref{fig:Egrowthb} shows that the growth rate reaches peaks between $t\sim 2$ to $t\sim 4$, depending on the seeded mode. The early transient is due to imperfect initial conditions which causes the particles to be accelerated or decelerated until they reach their equilibrium velocity. The time window around the peak growth rate is associated with the exponential growth of kinetic energy seen in figure \ref{fig:Egrowtha}, and is thus, closest to the dynamics predicted by the linear stability analysis. The growth rate in figure \ref{fig:Egrowthb} eventually drops, as the perturbation kinetic energy in \ref{fig:Egrowtha} saturates. The deviation from exponential growth correlates with the emergence of non-linear effects in the flow and number density fields as seen in figure \ref{fig:vol_frac_m246}.
}

{\color{revision2}
In order to quantify the contribution of a specific azimuthal mode $m$ to the overall dynamics, we compute the kinetic energy associated with this mode using
\begin{equation}
    E_m=\iint\frac{1}{2} \overline{\bm{\hat{u}}_m(r,t)} \cdot {\bm{\hat{u}}_m(r,t)} dS = \int_0^{\infty}\frac{1}{2} \overline{\bm{\hat{u}}_m(r,t)} \cdot {\bm{\hat{u}}_m(r,t)} 2\pi r dr
\end{equation}
where $\hat{\bm{u}}_m$ is the Fourier amplitude and reads
\begin{equation}
    \bm{\hat{u}}_m(r,t)=\frac{1}{2\pi} \int_0^{2\pi} (\bm{u}-\bm{u}_f^b) e^{-i m \theta} d\theta.
\end{equation}
}
{\color{revision2}Figure \ref{fig:mEgrowth} shows the time evolution of the total perturbation kinetic energy along side the contributions of  modes 2 to 5 for simulation cases A, B, C, and D. While the total perturbation energy is the sum of the energies of all Fourier modes, the mode initially seeded (i.e., $m=2$ for case A, $m=3$ for case B,  and so on) accounts for most of the perturbation energy during the early evolution $t\lesssim 4$. Thus, the growth rate computed from the total perturbation energy, shown in figure \ref{fig:Egrowthb}, matches the growth rate of the seeded mode early on. This is further evidenced in figure \ref{fig:mEgrowth_2} showing the evolution of the kinetic energy and growth rate associated with the seeded mode. The latter peaks between $t=2$ and $3$ depending on the wavenumber, at a level sensitively close to the peak growth rate of the total perturbation in figure \ref{fig:Egrowthb}.
At later times, the seeded mode is no longer dominant as the energy of other modes rises to a comparable level. Further, as time progresses, figure \ref{fig:mEgrowth} shows that the energy of several low-wavenumber modes drops, which corresponds to negative growth rates, also seen in figure \ref{fig:mEgrowth2}. This indicates a non-linear energy transfer from low to higher wavenumber modes.}

{\color{revision} Figure \ref{fig:compare} shows comparisons between growth rates predicted by the linear stability analysis and those obtained from Euler-Lagrange simulations (equation (\ref{eq:growth_rate})). {\color{revision2} From the latter, we report the total perturbation's peak growth rate and average value in a time window of size $\Delta t =1$ centered on the peak (see figure \ref{fig:Egrowthb}). We also report the peak growth rate of the seeded mode (see figure \ref{fig:mEgrowth2}).}
The difference between these values represents a confidence interval for the comparison of linear stability analysis and Euler-Lagrange simulations.
Figure \ref{fig:compare_a} and \ref{fig:compare_b} show the results at various mode number $m=$2 -- 6 for Atwood numbers $\mathrm{At}=1/3$ and 1/5, respectively. Both Euler-Lagrange simulations and linear stability analysis capture the growth rate increase  with increasing mode number $m$. The quantitative agreement is good given the different methodologies and underlying assumptions. The growth rates at various Atwood number for mode $m=3$ are shown in figure \ref{fig:compare_c}. Here the Euler-Lagrange simulations and linear stability analysis reproduce the same trend, that is the instability growth rate increases with Atwood number $\Atw$. Further, the peak growth rates obtained by the fully non-linear numerical simulation has excellent agreement with linear stability predictions.}

While the growth rate obtained from  numerical simulations and linear stability analysis are within close quantitative agreement, some of the differences can be related to the discretization of the eigenmodes and Lagrangian representation of the particles in the numerical simulations. Whereas eigenmodes derived from linear stability analysis require strict discontinuities in number density and vorticity, these effects are necessarily smoothed out when discretized on a mesh. Further results from the linear stability analysis are derived on the premise that viscous effects and discrete particle effects are negligible. However, these effects cannot be ruled out in the Euler-Lagrange simulations where viscosity still plays a role despite the large $\Rey_\Gamma=5000$ considered, and Lagrangian particles trigger high wavenumber modes. Despite the differences between the linear stability analysis and the Euler-Lagrange model, the agreement between the two approaches is satisfactory.

\section{Conclusion} \label{sec:conclusion}

In this work, we have shown that introducing heavy particles in the core of a vortex causes the breakdown of the flow structure due to an instability activated by two-way coupling, i.e., momentum exchange between the two phases. If particle feedback is neglected (one-way coupling), no instability is observed: the vortex retains its coherent axisymmetric structure while inertial particles are slowly expelled  outward forming a ring of clustered particles around the core. The inclusion of two-way coupling breaks the axisymmetry and gives rise to azimuthal perturbations in both number density and vorticity fields. As these perturbations develop, they turn into spiraling arms of concentrated particles emanating out of the core while regions of particle-free flow are sucked inward. The vorticity field displays similar pattern which cause the breakdown of the initial Rankine structure. Remarkably, this breakdown occurs even for inertia-less particles, i.e., zero-Stokes number, provided that the  mass loading is not vanishingly small. Particle inertia plays a dual role, in that, it destabilizes low wavenumber modes, but reduces the growth rate of high wavenumber ones.

The mechanisms driving the instability are characterized using linear stability analysis of a base state consisting of a Rankine vortex with core radius $r_c$ and a circular patch of particles with radius $r_p$. To describe the two-phase flow, we use an inviscid Two-Fluid model, and assume that the particles are weakly inertial, i.e., their Stokes number $\Sto$ is small but not zero. This assumption has two merits. First, it allows us to approximate the particle velocity field in terms of the carrier fluid velocity field and an inertial correction, thus, reducing the number of equations to solve. Second, it justifies a quasi-steady approach in which the time-dependence of the base state is ignored enabling us to carry out a classic linear stability analysis. This step is justified by the fact that variations of the base state have a timescale $\tau_c=\tau_f/\Sto$ which becomes very large in the limit $\Sto\ll 1$ \citep{shuaiAcceleratedDecayLambOseen2022}. 

Analysis of growth rates obtained from linear stability analysis reveals the existence of unstable modes regardless of the value of the Stokes number $\Sto$ and as long as the mass loading $M\neq 0$. Generally, the most unstable configuration corresponds to matching particle patch and vortex core radii, i.e., $r_c=r_p$. For $\Sto=0$ particles, the problem becomes analogous to a vortex with discontinuous radial density stratification described by \citet{dixitStabilityVortexRadial2011}. Modes with wavenumber $m$ above a critical wavenumber $m_\mathrm{cr}= (1+M)^2/((2+M)M)$ are unstable and have a growth rate $\Re\{\sigma\}\sim \sqrt{m \Atw}$ when $m\gg 1$, and where the Atwood number $\Atw=M/(2+M)$ provides a measure of the relative magnitude of the density jump between inner (particle-rich) and outer (particle-free) regions ($0\leq \Atw\leq 1$). When the particles are inertial ($\Sto\neq 0$) all modes become unstable regardless $\Sto$ and $M\neq 0$, even those with $m<m_\mathrm{cr}$. However, the growth rate of modes $m\geq m_\mathrm{cr}$ is reduced compared to the case $\Sto=0$. For large wavenumber modes $m\gg 1$, the growth rate approaches $\Re\{\sigma\}\sim \sqrt{m \Atw}(1-2\Sto\sqrt{m \Atw}) +O(\Sto^2)$.
For both inertial and non-inertial particles, the tendency of growth rates to increase with wavenumber $m$ suggests that modes that naturally emerge in experiments and simulations shall have a high-wavenumber. This mode is likely determined by a balance between viscous effects and preferential concentration, although this cannot be ascertained as we have not considered viscous effects in our linear stability analysis.

In addition to the linear stability analysis, Eulerian-Lagrangian simulations are carried out at $\Sto=0.002$, $\Rey=5000$, and mass loading $M=$ 0.4 -- 1. The simulations are initialized with the base state superimposed with eigenmodes found from linear stability analysis. The simulations allow us to explore the transition from linear to nonlinear regimes. During the early evolution of the perturbations, growth rates computed from the perturbation kinetic energy in the Eulerian-Lagrangian simulations show excellent agreement with growth rates predicted by the linear stability analysis for modes $m=$ 2 -- 6 and mass loadings $M=$ 0.4 -- 1 (or $\Atw=$ 1/6 -- 1/3). When the flow enters a nonlinear stage, we observe the emergence of a high-wave number mode with $m\sim 24$ -- 26. This mode develops into spiraling arms of number density and vorticity that ultimately cause the breakdown of the initial base state.

One must note that the present study has a key difference from prior investigations of dusty Kolmogorov flow and Rayleigh-Taylor turbulence -- a particle-free vortex is always stable. Our study emphasizes that the disperse phase can also be the source of a new instability besides modifying an existing one. The novel instability for a dusty vortex identified here highlights how the feedback force from a disperse phase can induce a breakdown of an otherwise resilient vortical structure.

\vspace{2ex}
\noindent{\bf Acknowledgements}. The authors acknowledge support from the US National Science Foundation (award \#2148710, CBET-PMP), and from IIT Madras for the ``Geophysical Flows Lab'' research initiative under the Institute of Eminence framework.

\vspace{2ex}
\noindent{\bf Declaration of Interests}. The authors report no conflict of interest.

\bibliography{jfm,references_houssem}

\begin{thebibliography}{50}
\expandafter\ifx\csname natexlab\endcsname\relax\def\natexlab#1{#1}\fi
\def\au#1{#1} \def\ed#1{#1} \def\yr#1{#1}\def\at#1{#1}\def\jt#1{\textit{#1}}
  \def\bt#1{#1}\def\bvol#1{\textbf{#1}} \def\vol#1{#1} \def\pg#1{#1}
  \def\publ#1{#1}\def\arxiv#1{#1}\def\org#1{#1}\def\st#1{\textit{#1}}

\bibitem[Anderson \& Jackson(1967)]{andersonFluidMechanicalDescription1967}
{\sc \au{Anderson, T.~B.} \& \au{Jackson, R.}} \yr{1967}  \at{Fluid
  {{Mechanical Description}} of {{Fluidized Beds}}. {{Equations}} of
  {{Motion}}}.  \jt{Industrial \& Engineering Chemistry Fundamentals}
  \bvol{6}~(4),  \pg{527--539}.

\bibitem[Balachandar \& Eaton(2010)]{balachandar2010turbulent}
{\sc \au{Balachandar, S.} \& \au{Eaton, J.~K.}} \yr{2010}  \at{Turbulent
  dispersed multiphase flow}.  \jt{Annual review of fluid mechanics}
  \bvol{42},  \pg{111--133}.

\bibitem[Balmforth {\em et~al.\/}(2012)Balmforth, Roy \&
  Caulfield]{balmforth2012dynamics}
{\sc \au{Balmforth, N.J.}, \au{Roy, A} \& \au{Caulfield, C.P.}} \yr{2012}
  \at{Dynamics of vorticity defects in stratified shear flow}.  \jt{Journal of
  Fluid Mechanics}  \bvol{694},  \pg{292--331}.

\bibitem[Batchelor \&
  Nitsche(1991)]{batchelorInstabilityStationaryUnbounded1991}
{\sc \au{Batchelor, G.~K.} \& \au{Nitsche, J.~M.}} \yr{1991}  \at{Instability
  of stationary unbounded stratified fluid}.  \jt{Journal of Fluid Mechanics}
  \bvol{227},  \pg{357--391}.

\bibitem[Bluestein {\em et~al.\/}(2003)Bluestein, W., Bell, Weiss \&
  Pazmany]{bluesteinMobileDopplerRadar2003}
{\sc \au{Bluestein, H.~B.}, \au{W., Lee}, \au{Bell, M.}, \au{Weiss, C.~C.} \&
  \au{Pazmany, A.~L.}} \yr{2003}  \at{Mobile {{Doppler Radar Observations}} of
  a {{Tornado}} in a {{Supercell}} near {{Bassett}}, {{Nebraska}}, on 5
  {{June}} 1999. {{Part II}}: {{Tornado}}-{{Vortex Structure}}}.  \jt{Monthly
  Weather Review}  \bvol{131}~(12),  \pg{2968--2984}.

\bibitem[Bluestein {\em et~al.\/}(2004)Bluestein, Weiss \&
  Pazmany]{bluesteinDopplerRadarObservations2004}
{\sc \au{Bluestein, H.~B.}, \au{Weiss, C.~C.} \& \au{Pazmany, A.~L.}} \yr{2004}
   \at{Doppler {{Radar Observations}} of {{Dust Devils}} in {{Texas}}}.
  \jt{Monthly Weather Review}  \bvol{132}~(1),  \pg{209--224}.

\bibitem[Broadbent \& Moore(1979)]{broadbent1979acoustic}
{\sc \au{Broadbent, E.~G.} \& \au{Moore, D.~W.}} \yr{1979}  \at{Acoustic
  destablilization of vortices}.  \jt{Philosophical Transactions of the Royal
  Society of London. Series A, Mathematical and Physical Sciences}
  \bvol{290}~(1372),  \pg{353--371}.

\bibitem[Brown {\em et~al.\/}(2005)Brown, Flickinger, Forren, Schultz \&
  {al}]{brownImprovedDetectionSevere2005}
{\sc \au{Brown, R.~A.}, \au{Flickinger, B.~A.}, \au{Forren, E.}, \au{Schultz,
  D.~M.} \& \au{{al}, et}} \yr{2005}  \at{Improved {{Detection}} of {{Severe
  Storms Using Experimental Fine}}-{{Resolution WSR}}-{{88D Measurements}}}.
  \jt{Weather and Forecasting}  \bvol{20}~(1),  \pg{3--10,12--14}.

\bibitem[Capecelatro \& Desjardins(2013)]{capecelatroEulerLagrangeStrategy2013}
{\sc \au{Capecelatro, J.} \& \au{Desjardins, O.}} \yr{2013}  \at{An
  {{Euler}}\textendash{{Lagrange}} strategy for simulating particle-laden
  flows}.  \jt{Journal of Computational Physics}  \bvol{238},  \pg{1--31}.

\bibitem[Carpenter {\em et~al.\/}(2011)Carpenter, Tedford, Heifetz \&
  Lawrence]{carpenter2011instability}
{\sc \au{Carpenter, Jeffrey~R}, \au{Tedford, Edmund~W}, \au{Heifetz, Eyal} \&
  \au{Lawrence, Gregory~A}} \yr{2011}  \at{Instability in stratified shear
  flow: Review of a physical interpretation based on interacting waves}.
  \jt{Applied Mechanics Reviews}  \bvol{64}~(6).

\bibitem[Dixit \&
  Govindarajan(2010)]{dixitVortexinducedInstabilitiesAccelerated2010a}
{\sc \au{Dixit, H.~N} \& \au{Govindarajan, R.}} \yr{2010}  \at{Vortex-induced
  instabilities and accelerated collapse due to inertial effects of density
  stratification}.  \jt{Journal of Fluid Mechanics}  \bvol{646},
  \pg{415--439}.

\bibitem[Dixit \& Govindarajan(2011)]{dixitStabilityVortexRadial2011}
{\sc \au{Dixit, H.~N.} \& \au{Govindarajan, R.}} \yr{2011}  \at{Stability of a
  vortex in radial density stratification: Role of wave interactions}.
  \jt{Journal of Fluid Mechanics}  \bvol{679},  \pg{582--615}.

\bibitem[Druzhinin(1994)]{druzhininConcentrationWavesFlow1994}
{\sc \au{Druzhinin, O.~A.}} \yr{1994}  \at{Concentration waves and flow
  modification in a particle-laden circular vortex}.  \jt{Physics of Fluids}
  \bvol{6}~(10),  \pg{3276--3284}.

\bibitem[Druzhinin(1995)]{druzhininTwowayInteractionTwodimensional1995}
{\sc \au{Druzhinin, O.~A.}} \yr{1995}  \at{On the two-way interaction in
  two-dimensional particle-laden flows: The accumulation of particles and flow
  modification}.  \jt{Journal of Fluid Mechanics}  \bvol{297},  \pg{49--76}.

\bibitem[Ferry \& Balachandar(2001)]{ferryFastEulerianMethod2001}
{\sc \au{Ferry, J.} \& \au{Balachandar, S.}} \yr{2001}  \at{A fast {{Eulerian}}
  method for disperse two-phase flow}.  \jt{International Journal of Multiphase
  Flow}  \bvol{27}~(7),  \pg{1199--1226}.

\bibitem[Ferry \& Balachandar(2002)]{ferryEquilibriumExpansionEulerian2002}
{\sc \au{Ferry, J.} \& \au{Balachandar, S.}} \yr{2002}  \at{Equilibrium
  expansion for the {{Eulerian}} velocity of small particles}.  \jt{Powder
  Technology}  \bvol{125}~(2\textendash 3),  \pg{131--139}.

\bibitem[Ford(1994)]{ford1994instability}
{\sc \au{Ford, R.}} \yr{1994}  \at{The instability of an axisymmetric vortex
  with monotonic potential vorticity in rotating shallow water}.  \jt{Journal
  of Fluid Mechanics}  \bvol{280},  \pg{303--334}.

\bibitem[Fung \& Kurzweg(1975)]{fungStabilitySwirlingFlows1975}
{\sc \au{Fung, Y.~T.} \& \au{Kurzweg, U.~H.}} \yr{1975}  \at{Stability of
  swirling flows with radius-dependent density}.  \jt{Journal of Fluid
  Mechanics}  \bvol{72}~(02),  \pg{243}.

\bibitem[Guazzelli \& Morris(2011)]{guazzelli2011physical}
{\sc \au{Guazzelli, E.} \& \au{Morris, J.~F}} \yr{2011} {\em A physical
  introduction to suspension dynamics\/}, ,  \vol{vol.~45}.  \publ{Cambridge
  University Press}.

\bibitem[Jackson(2000)]{jacksonDynamicsFluidizedParticles2000}
{\sc \au{Jackson, Roy}} \yr{2000} {\em The {{Dynamics}} of {{Fluidized
  Particles}}\/}.  \publ{{Cambridge University Press}}.

\bibitem[Joly {\em et~al.\/}(2005)Joly, Fontane \&
  Chassaing]{jolyRayleighTaylorInstability2005}
{\sc \au{Joly, L.}, \au{Fontane, J.} \& \au{Chassaing, P.}} \yr{2005}  \at{The
  {{Rayleigh}}\textendash{{Taylor}} instability of two-dimensional high-density
  vortices}.  \jt{Journal of Fluid Mechanics}  \bvol{537}~(-1),  \pg{415}.

\bibitem[Kasbaoui(2019)]{kasbaouiTurbulenceModulationSettling2019}
{\sc \au{Kasbaoui, M.~H.}} \yr{2019}  \at{Turbulence modulation by settling
  inertial aerosols in {{Eulerian-Eulerian}} and {{Eulerian-Lagrangian}}
  simulations of homogeneously sheared turbulence}.  \jt{Physical Review
  Fluids}  \bvol{4}~(12),  \pg{124308}.

\bibitem[Kasbaoui {\em et~al.\/}(2019{\natexlab{{\em a\/}}})Kasbaoui, Koch \&
  Desjardins]{kasbaouiClusteringEulerEuler2019}
{\sc \au{Kasbaoui, M.~H.}, \au{Koch, D.~L.} \& \au{Desjardins, O.}}
  \yr{2019{\natexlab{{\em a\/}}}}  \at{Clustering in
  {{Euler}}\textendash{{Euler}} and {{Euler}}\textendash{{Lagrange}}
  simulations of unbounded homogeneous particle-laden shear}.  \jt{Journal of
  Fluid Mechanics}  \bvol{859},  \pg{174--203}.

\bibitem[Kasbaoui {\em et~al.\/}(2019{\natexlab{{\em b\/}}})Kasbaoui, Koch \&
  Desjardins]{kasbaouiRapidDistortionTwoway2019}
{\sc \au{Kasbaoui, M.~H.}, \au{Koch, D.~L.} \& \au{Desjardins, O.}}
  \yr{2019{\natexlab{{\em b\/}}}}  \at{The rapid distortion of two-way coupled
  particle-laden turbulence}.  \jt{Journal of Fluid Mechanics}  \bvol{877},
  \pg{82--104}.

\bibitem[Kasbaoui {\em et~al.\/}(2015)Kasbaoui, Koch \&
  Subramanian]{kasbaouiPreferentialConcentrationDriven2015}
{\sc \au{Kasbaoui, M.~H.}, \au{Koch, D.~L.} \& \au{Subramanian,
  G.and~Desjardins, O.}} \yr{2015}  \at{Preferential concentration driven
  instability of sheared gas\textendash solid suspensions}.  \jt{Journal of
  Fluid Mechanics}  \bvol{770},  \pg{85--123}.

\bibitem[Lamb(1993)]{lambHydrodynamics1993}
{\sc \au{Lamb, H.}} \yr{1993} {\em Hydrodynamics\/}.  \publ{{Cambridge; New
  York}: {Cambridge University Press}}.

\bibitem[Le~Diz{\`e}s \& Billant(2009)]{le2009radiative}
{\sc \au{Le~Diz{\`e}s, S.} \& \au{Billant, P.}} \yr{2009}  \at{Radiative
  instability in stratified vortices}.  \jt{Physics of Fluids}  \bvol{21}~(9),
  \pg{096602}.

\bibitem[Magnani {\em et~al.\/}(2021)Magnani, Musacchio \&
  Boffetta]{magnani2021inertial}
{\sc \au{Magnani, M.}, \au{Musacchio, S.} \& \au{Boffetta, G.}} \yr{2021}
  \at{Inertial effects in dusty rayleigh--taylor turbulence}.  \jt{Journal of
  Fluid Mechanics}  \bvol{926}.

\bibitem[Marble(1970)]{marbleDynamicsDustyGases1970}
{\sc \au{Marble, F~E}} \yr{1970}  \at{Dynamics of {{Dusty Gases}}}.  \jt{Annual
  Review of Fluid Mechanics}  \bvol{2}~(1),  \pg{397--446}.

\bibitem[Marshall(2005)]{marshallParticleDispersionTurbulent2005b}
{\sc \au{Marshall, J.~S.}} \yr{2005}  \at{Particle dispersion in a turbulent
  vortex core}.  \jt{Physics of Fluids}  \bvol{17}~(2),  \pg{025104}.

\bibitem[Maxey(1987)]{maxeyGravitationalSettlingAerosol1987}
{\sc \au{Maxey, M.~R.}} \yr{1987}  \at{The gravitational settling of aerosol
  particles in homogeneous turbulence and random flow fields}.  \jt{Journal of
  Fluid Mechanics}  \bvol{174},  \pg{441--465}.

\bibitem[Maxey \& Riley(1983)]{maxeyEquationMotionSmall1983}
{\sc \au{Maxey, M.~R.} \& \au{Riley, J.~J.}} \yr{1983}  \at{Equation of motion
  for a small rigid sphere in a nonuniform flow}.  \jt{The Physics of Fluids}
  \bvol{26}~(4),  \pg{883--889}.

\bibitem[Michalke \& Timme(1967)]{michalkeInviscidInstabilityCertain1967}
{\sc \au{Michalke, Alfons} \& \au{Timme, Adalbert}} \yr{1967}  \at{On the
  inviscid instability of certain two-dimensional vortex-type flows}.
  \jt{Journal of Fluid Mechanics}  \bvol{29}~(4),  \pg{647--666}.

\bibitem[Paoli \& Shariff(2016)]{paoliContrailModelingSimulation2016}
{\sc \au{Paoli, R.} \& \au{Shariff, K.}} \yr{2016}  \at{Contrail {{Modeling}}
  and {{Simulation}}}.  \jt{Annual Review of Fluid Mechanics}  \bvol{48}~(1),
  \pg{393--427}.

\bibitem[Paoli {\em et~al.\/}(2008)Paoli, Vancassel, Garnier \&
  Mirabel]{paoliLargeeddySimulationTurbulent2008}
{\sc \au{Paoli, R.}, \au{Vancassel, X.}, \au{Garnier, F.} \& \au{Mirabel, P.}}
  \yr{2008}  \at{Large-eddy simulation of a turbulent jet and a vortex sheet
  interaction: Particle formation and evolution in the near field of an
  aircraft wake}.  \jt{Meteorologische Zeitschrift}  \bvol{17}~(2),
  \pg{131--144}.

\bibitem[Rani \& Balachandar(2003)]{raniEvaluationEquilibriumEulerian2003}
{\sc \au{Rani, S.~L.} \& \au{Balachandar, S.}} \yr{2003}  \at{Evaluation of the
  equilibrium {{Eulerian}} approach for the evolution of particle concentration
  in isotropic turbulence}.  \jt{International Journal of Multiphase Flow}
  \bvol{29}~(12),  \pg{1793--1816}.

\bibitem[Ravichandran \&
  Govindarajan(2015)]{ravichandranCausticsClusteringVicinity2015a}
{\sc \au{Ravichandran, S.} \& \au{Govindarajan, R.}} \yr{2015}  \at{Caustics
  and clustering in the vicinity of a vortex}.  \jt{Physics of Fluids}
  \bvol{27}~(3),  \pg{033305}.

\bibitem[Roy {\em et~al.\/}(2022)Roy, Garg, Reddy \&
  Subramanian]{roy2022inertio}
{\sc \au{Roy, Anubhab}, \au{Garg, Piyush}, \au{Reddy, Jumpal~Shashikiran} \&
  \au{Subramanian, Ganesh}} \yr{2022}  \at{Inertio--elastic instability of a
  vortex column}.  \jt{Journal of Fluid Mechanics}  \bvol{937}.

\bibitem[Roy \& Subramanian(2014)]{roy2014linearized}
{\sc \au{Roy, A.} \& \au{Subramanian, G.}} \yr{2014}  \at{Linearized
  oscillations of a vortex column: the singular eigenfunctions}.  \jt{Journal
  of Fluid Mechanics}  \bvol{741},  \pg{404--460}.

\bibitem[Saffman(1995)]{saffman1995vortex}
{\sc \au{Saffman, P.~G.}} \yr{1995} {\em Vortex dynamics\/}.  \publ{Cambridge
  university press}.

\bibitem[Schecter \& Montgomery(2004)]{schecter2004damping}
{\sc \au{Schecter, D.~A.} \& \au{Montgomery, M.~T.}} \yr{2004}  \at{Damping and
  pumping of a vortex rossby wave in a monotonic cyclone: critical layer
  stirring versus inertia--buoyancy wave emission}.  \jt{Physics of Fluids}
  \bvol{16}~(5),  \pg{1334--1348}.

\bibitem[Shuai \& Kasbaoui(2022)]{shuaiAcceleratedDecayLambOseen2022}
{\sc \au{Shuai, S.} \& \au{Kasbaoui, M.~Houssem}} \yr{2022}  \at{Accelerated
  decay of a {{Lamb-Oseen}} vortex tube laden with inertial particles in
  {{Eulerian-Lagrangian}} simulations}.  \jt{Journal of Fluid Mechanics}
  \bvol{936}.

\bibitem[Sipp {\em et~al.\/}(2005{\natexlab{{\em a\/}}})Sipp, Fabre, Michelin
  \& Jacquin]{sippStabilityVortexHeavy2005a}
{\sc \au{Sipp, D.}, \au{Fabre, D.}, \au{Michelin, S.} \& \au{Jacquin, L.}}
  \yr{2005{\natexlab{{\em a\/}}}}  \at{Stability of a vortex with a heavy
  core}.  \jt{Journal of Fluid Mechanics}  \bvol{526},  \pg{67--76}.

\bibitem[Sipp {\em et~al.\/}(2005{\natexlab{{\em b\/}}})Sipp, Fabre, Michelin
  \& Jacquin]{sippStabilityVortexHeavy2005}
{\sc \au{Sipp, D.}, \au{Fabre, D.}, \au{Michelin, S.} \& \au{Jacquin, L.}}
  \yr{2005{\natexlab{{\em b\/}}}}  \at{Stability of a vortex with a heavy
  core}.  \jt{Journal of Fluid Mechanics}  \bvol{526},  \pg{67--76}.

\bibitem[Sozou \& Lighthill(1987)]{sozouAdiabaticPerturbationsUnbounded1987}
{\sc \au{Sozou, C.} \& \au{Lighthill, M.~J.}} \yr{1987}  \at{Adiabatic
  perturbations in an unbounded {{Rankine}} vortex}.  \jt{Proceedings of the
  Royal Society of London. A. Mathematical and Physical Sciences}
  \bvol{411}~(1840),  \pg{207--224}.

\bibitem[Sozza {\em et~al.\/}(2022)Sozza, Cencini, Musacchio \&
  Boffetta]{sozza2022instability}
{\sc \au{Sozza, A.}, \au{Cencini, M.}, \au{Musacchio, S.} \& \au{Boffetta, G.}}
  \yr{2022}  \at{Instability of a dusty kolmogorov flow}.  \jt{Journal of Fluid
  Mechanics}  \bvol{931}.

\bibitem[Squires \& Eaton(1991)]{squiresPreferentialConcentrationParticles1991}
{\sc \au{Squires, K.~D.} \& \au{Eaton, J.~K.}} \yr{1991}  \at{Preferential
  concentration of particles by turbulence}.  \jt{Physics of Fluids A: Fluid
  Dynamics}  \bvol{3}~(5),  \pg{1169--1178}.

\bibitem[Thomson(1880)]{thomsonXXIVVibrationsColumnar1880}
{\sc \au{Thomson, W.}} \yr{1880}  \at{{{XXIV}}. {{Vibrations}} of a columnar
  vortex}.  \jt{The London, Edinburgh, and Dublin Philosophical Magazine and
  Journal of Science}  \bvol{10}~(61),  \pg{155--168}.

\bibitem[Trefethen(2000)]{trefethen2000spectral}
{\sc \au{Trefethen, L.~N.}} \yr{2000} {\em Spectral methods in MATLAB\/}.
  \publ{SIAM}.

\bibitem[Wang \& Maxey(1993)]{wangSettlingVelocityConcentration1993}
{\sc \au{Wang, L.} \& \au{Maxey, M.~R.}} \yr{1993}  \at{Settling velocity and
  concentration distribution of heavy particles in homogeneous isotropic
  turbulence}.  \jt{Journal of Fluid Mechanics}  \bvol{256},  \pg{27--68}.

\end{thebibliography}
\end{document}